\documentclass[12pt]{article}
\usepackage{amsmath}
\usepackage{graphicx,psfrag,epsf}
\usepackage{enumerate}
\usepackage{natbib}
\usepackage{url} % not crucial - just used below for the URL 
\usepackage{caption}
\usepackage{subcaption}
\usepackage{dcolumn}
\usepackage{lscape}

\usepackage{bm}
\usepackage{tikz}
\usepackage{mathtools}
\newcommand{\indep}{\rotatebox[origin=c]{90}{$\models$}}
\usepackage[flushleft]{threeparttable}

\newcommand{\blind}{0}

\begin{document}

\def\spacingset#1{\renewcommand{\baselinestretch}%
{#1}\small\normalsize} \spacingset{1}

%%%%%%%%%%%%%%%%%%%%%%%%%%%%%%%%%%%%%%%%%%%%%%%%%%%%%%%%%%%%%%%%%%%%%%%%%%%%%%

\if0\blind
{
  \title{\bf Efficient, Doubly Robust Estimation of the Effect of Dose Switching for Switchers in a Randomised Clinical Trial}
  \author{Kelly Van Lancker\\
    Department of Applied Mathematics, Computer Science and Statistics,\\ Ghent University, Ghent, Belgium\\
    An Vandebosch \\
    Janssen R\&D,\\ a division of Janssen Pharmaceutica NV, Beerse, Belgium\\
    and \\
	Stijn Vansteelandt\\
Department of Applied Mathematics, Computer Science and Statistics,\\ Ghent University, Ghent, Belgium\\
Department of Medical Statistics,\\ London School of Hygiene and Tropical Medicine, London, United Kingdom\\}
  \maketitle
} \fi

\if1\blind
{
  \bigskip
  \bigskip
  \bigskip
  \begin{center}
    {\LARGE\bf Title}
\end{center}
  \medskip
} \fi

\bigskip
\begin{abstract}
Motivated by a clinical trial conducted by Janssen Pharmaceuticals in which a flexible dosing regimen is compared to placebo, we evaluate how switchers in the treatment arm (i.e., patients who were switched to the higher dose) would have fared had they been kept on the low dose. This in order to understand whether flexible dosing is potentially beneficial for them.  Simply comparing these patients' responses with those of patients who stayed on the low dose is unsatisfactory because the latter patients are usually in a better health condition.  Because the available information in the considered trial is too scarce to enable a reliable adjustment, we will instead transport data from a fixed dosing trial that has been conducted concurrently on the same target, albeit not in an identical patient population. In particular, we will propose an estimator which relies on an outcome model and a propensity score model for the association between study and patient characteristics. The proposed estimator is asymptotically unbiased if at least one of both models is correctly specified, and efficient (under the model defined by the restrictions on the propensity score) when both models are correctly specified. We show that the proposed method for using results from an external study is generically applicable in studies where a classical confounding adjustment is not possible due to positivity violation (e.g., studies where switching takes place in a deterministic manner).
Monte Carlo simulations and application to the motivating study demonstrate adequate performance.
\end{abstract}

\noindent%
\vfill

\newpage
\spacingset{1.45} % DON'T change the spacing!
\section{Introduction}
In drug development, it is generally recommended to achieve at least two positive efficacy trials before applying for health authority approval \citep{FDA}. Therefore the development program typically consists of multiple phase $3$ studies, which are not necessarily identical or are not necessarily performed in the same population.
%To achieve the two (or more) positive efficacy trials required for Food and Drug Administration (FDA) marketing approval,
%To achieve the requirement of the Food and Drug Administration (FDA) for grant approval, i.e.\,two or more positive efficacy trials in a defined patient population,
%pharmaceutical companies must at least conduct two phase $3$ trials- but typically many more to account for the possibility of negative studies \citep{FDA}.
In patients with a chronic condition (e.g., Schizophrenia in neurosience), for example, two of these trials are typically a fixed and flexible dosing trial \citep[e.g.,][]{Pollack1998, EMA_Abilify, EMA_invega, Lundbeck}.
%(e.g.\,, Invega, Xeplion, \dots). 
This is often done in practice because one can learn different features from these studies; a fixed dosing trial 
evaluates the differences in safety, tolerability and efficacy between doses, while a flexible dosing study more accurately reflects real-life dosing strategies \citep{Lipkovich2011}.  

The use of both types of trials appeared for instance in a study program  at Janssen Pharmaceuticals,
% for treatment-resistant depression, 
which motivated this research. This program included a fixed dosing trial and a flexible dosing trial with protocols that were similar in target patient population (I/E criteria), primary endpoint and treatment effect measure for the primary endpoint.
In the fixed dosing trial a fixed low dose, a fixed high dose and a (fixed) placebo were compared. In the flexible dosing trial a (flexible) placebo was compared to a flexible dose, where patients start with the low dose but are able to switch to the high dose (and vice versa) depending on the investigator's decision using a pre-specified deterministic rule  based on efficacy and tolerability. 
Results from the flexible dosing trial showed that the flexible dose demonstrated a clinically meaningful and statistically significant improvement in clinical outcome
%depressive symptoms 
compared to (flexible) placebo. 
In the fixed dosing trial, 
%numerically lower depressive symptoms scores 
numerically better clinical outcomes were observed in both dose groups compared to placebo. However, no dose-response relationship was observed.
%However, in the fixed dosing trial, a treatment effect was observed at both doses, but no dose-response relationship.
This raised the question whether flexible dosing is potentially beneficial for switchers in the treatment arm (i.e., patients who were switched to the higher dose). 
This raised the question whether the high dose was potentially beneficial for a certain subgroup of patients - in particular, whether flexible dosing is potentially beneficial for switchers in the flexible dosing arm (i.e., patients who were switched to the higher dose). 
We therefore aim to estimate the treatment effect in these switchers, which
expresses how different the outcome would have been for them, had they been kept on the low dose (i.e.\,, not switched).
Simply comparing these patients' responses with those of patients in the flexible dosing arm of the flexible dosing study who stayed on the low dose is unsatisfactory because these groups are usually not comparable (e.g.\,, the latter patients are usually in a better health condition).
Additionally, the deterministic rule for switching complicates a simple adjustment for confounding such as inverse probability weighting due to positivity violation.
Because the available information in the considered trial is too scarce to enable a relaible adjustment, we will instead use the information contained in the low dose arm of the fixed dosing trial.
%Moreover, the available information in the flexible dosing trial is too scarce to enable a reliable adjustment to estimate the treatment effect in the switchers since no patients have been assigned to the fixed low dose in the flexible dosing trial.
%In view of this, we will transport data from the low dose arm of the fixed dosing trial to the flexible dosing trial.
Although the fixed and flexible dosing trials have been conducted concurrently on the same (proposed) treatment, 
%for treatment-resistant depression, 
the patient populations may differ with respect to the distribution of covariates. 
%Because of such differences, the average of the observed outcome under the fixed low dose in the fixed dosing trial may differ from the average outcome we would observe in the flexible dosing trial if they had been assigned to the fixed low dose. This makes it impossible to directly compare the results of the patients in the flexible dosing arm of the flexible dosing trial to the results of the patients in the low dose arm of the fixed dosing trial.
Because of such differences, it is also impossible to directly compare the results of the patients in the flexible dosing arm of the flexible dosing trial to the results of the patients in the low dose arm of the fixed dosing trial.

In view of this, we will transport data from the low dose arm of the fixed dosing trial to the flexible dosing trial by using similar statistical techniques as for transporting inferences from trial participants to new target populations \citep{Zhang2016, Dahabreh2018}. We formulate the statistical problem in terms of potential outcomes \citep{Rubin1974} and discuss the plausibility of the assumptions required to transport the mean of the potential outcomes under
a given treatment regimen from one trial to another. To avoid undue reliance on modelling assumptions, we propose a doubly robust estimator. This estimator relies on an outcome model and a selection (i.e.\,, propensity score) model for the association between study and patient characteristics, but only requires at least one of them to be correctly specified \citep{Robins1994, Robins1995, Tsiatis2006}. 
The proposed method for using results from an external study is generically applicable and creates possibilities in studies where a classical confounding adjustment is not possible due to positivity violation (e.g,. studies where switching takes place in a deterministic manner).
We moreover assess the finite-sample performance of the method in simulation studies and support the proposal with an application of the methods to the motivating study. Lastly, we discuss practical considerations regarding trial designs and problems that can arise when applying transportability methods.
%variable selection and model specification. 
%Although the fixed results show that the low dose is more effective than the high dose (and thus that they will probably only launch the low dose on the market), it may be that the high dose is more effective for a certain group of patients, namely those who switched to the higher dose in the flexible dosing study. It is really important to investigate this because if they have more benefit from the higher dose it is important to not only launch (bring on the market) the low drug but also the high drug.

\section{Setting and Estimation} \label{sec:est}
\subsection{Trial Design}\label{setting1}
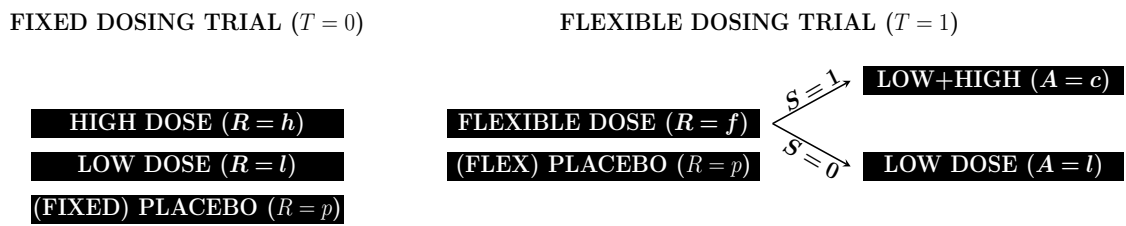
\begin{figure}[h!]
	\resizebox{15cm}{3cm}{%
		\begin{tikzpicture}
		\node[white] at(10.25,2)  {\textbf{(FLEXIBLE) PLACEBO ($\bm{A=p}$)}};
		
		\node[black] at(4.25,2.5) { \textbf{FIXED DOSING TRIAL ($T=0$)}};
		\fill[] (1.25,1) rectangle (7.25, 0.5);
		\fill[black] (1.25,-0.25) rectangle (7.25, 0.25);
		\fill[black] (1.25,-0.5) rectangle (7.25, -1);
		\node[white] at(4.25,0.75) {\textbf{HIGH DOSE ($\bm{R=h}$)}};
		\node[white] at(4.25,0) {\textbf{LOW DOSE ($\bm{R=l}$)}};
		\node[white] at(4.25,-0.75) {\textbf{(FIXED) PLACEBO ($R=p$)} 
		};
		
		\node[black] at(15.25,2.5) { \textbf{FLEXIBLE DOSING TRIAL ($T=1$)}};
		\fill[black] (9.25,1) rectangle (15.25, 0.5);
		\fill[black] (9.25,-0.25) rectangle (15.25, 0.25);
		\node[white] at(12.25,0.75) {\textbf{FLEXIBLE DOSE ($\bm{R=f}$)}};
		\node[white] at(12.25,0) {\textbf{(FLEX) PLACEBO ($R=p$)}
		};

		\draw[black, thick, -stealth] (15.5,0.75) -- (17,1.5);
		\draw[black, thick, -stealth] (15.5,0.75) -- (17,0);
		
		\fill[black] (17.25,0.25) rectangle (22.25,-0.25);
		\node[white] at(19.75,0) {\textbf{LOW DOSE ($\bm{A=l}$)}};
		\fill[black] (17.25,1.25) rectangle (22.25, 1.75);
		\node[white] at(19.75,1.5) {\textbf{LOW+HIGH ($\bm{A=c}$)}};
		
		\node[rotate=28] at(16.25,1.35) {$\bm{S=1}$};
		\node[rotate=-28] at(16.25,0.15) {$\bm{S=0}$};
		
		\end{tikzpicture}
	}
	\caption{\label{fig:design}Trial designs}
\end{figure}

The observed data consist of independent and identically distributed observations $\{(\bm{Z}_i, Y_i, T_i, R_i): i=1, \dots, n\}$ on measured baseline covariates $\bm{Z}$, outcome $Y$, a trial indicator $T$ that takes the value $1$ or $0$ for subjects assigned to respectively the flexible and fixed dosing trial and the randomized treatment $R$ which is coded $p$ for patients randomized to placebo, $h$ or $l$ for patients in the fixed dosing trial randomized to respectively the fixed high and fixed low dose, and $f$ for patients in the flexible dosing trial randomized to the flexible dose (see Figure \ref{fig:design}). 
Depending on their response, patients assigned to the flexible dose either stay on the low dose or switch from the low dose to the high dose (or vice versa). As shown in Figure \ref{fig:design}, the actual assigned treatment will be denoted as $A=l$ for the non-switchers ($S=0$) since they stay assigned to the low dose and $A=c$ for the switchers ($S=1$) since they are assigned to a combination of the low and high dose. For the other patients, we do not distinguish between the randomized and assigned treatment ($A=R$). 
%\subsection{Research Question in Causal Framework}
\subsection{Proposal}\label{sec:proposal}
Let us write $Y^a$ to represent the counterfactual outcome
under treatment $A=a$ ($a \in \{l, h, c, p\}$). In particular, $Y^a$ is the outcome which would have been seen for given subject had (s)he been assigned to dose $a$. 
%In particular, each subject has a four-dimensional vector of counterfactual outcomes ($Y^l$, $Y^h$, $Y^c$, $Y^p$) but only one of them is observed and the others are counterfactual (``contrary to the facts''). For example, if $B = l$ then $Y^l$ is observed, while $Y^h$, $Y^c$ and $Y^p$ are not.
To investigate whether flexible dosing is potentially beneficial (in terms of treatment effect compared to the low dose) for switchers in the treatment arm of the flexible dosing study, we evaluate how switchers in the treatment arm would have fared had they been kept on the low dose. In particular, we compare the difference in means between the counterfactual outcomes under the flexible dose and the low dose for switchers in the treatment arm of the flexible dosing study. Since the observed outcome under the flexible dose for switchers is given by $Y^c$, we can write this difference in means as
\begin{align}\label{eq:quest}
E[Y^{c}-Y^{l}|T=1, R=f, S=1].
\end{align}	
Defining $Y^f$ as the counterfactual outcome under the flexible dose (not distinguishing between switchers and non-switchers), this can be rewritten as
\begin{align*}
&E[SY^{c}-SY^{l}|T=1, R=f]/P(S=1|T=1, R=f)\\ 
&=E[SY^{c}+(1-S)Y^{l}-SY^{l}-(1-S)Y^{l}|T=1, R=f]/\\
&\hspace{0.5cm}P(S=1|T=1, R=f)\\ 
&=E[Y^f-Y^{l}|T=1, R=f]/P(S=1|T=1, R=f).
%&=\left[E[Y^{\text{flex}}|S=1, A=1]-E[Y^{56}|S=1, A=1]\right]/P(R=0|S=1, A=1).
\end{align*}
by consecutively using the law of iterated expectation, adding and substracting the term $(1-S)Y^{l}$ and using that $SY^c+(1-S)Y^l=Y^f$.

The probability $P(S=1|T=1, R=f)$ can be simply estimated as the proportion of switchers in the flexible dosing arm. 

The first part of 
%the ATT\footnote{Note that this is an average treatment effect in the patients assigned to the flexible dose, rather than an average treatment effect (ATE).} effect
 $E[Y^f-Y^{l}|T=1, R=f]$, $\theta_1\coloneqq E[Y^{f}|T=1, R=f]$, can be directly estimated in the flexible dosing trial. In particular, since the flexible dosing trial is a randomized trial, an efficient (non-parametric) estimator $\hat{\theta}_1$ for $\theta_1$ which exploits the assumption that $R\indep \bm{Z}|T=1$ can be obtained in this trial as follows \citep{Tsiatis2006, ShuTan2018}
\begin{enumerate}
	\item regress $Y$ on $\bm{Z}$ among the patients on the flexible dose ($R=f$) using a canonical generalized linear working model for the conditional mean of $Y$: $E(Y|R=f, T=1, \bm{Z})=h(\bm{Z}, \bm{\eta}_0)$, where $h(\bm{Z}, \bm{\eta})$ is a known function, evaluated at a parameter $\bm{\eta}$ with unknown population value $\bm{\eta}_0$; e.g.\,, $h(\bm{Z}, \bm{\eta})=\eta_1+\bm{\eta}_2'\bm{Z}$ for a continuous outcome $Y$,
	\item and take the average of the predicted values over all patients in the flexible dosing trial.
\end{enumerate}
As a result of the simple randomisation in the flexible dosing trial, the estimator $\hat{\theta}_{1}$ has the appealing feature that misspecification of the outcome model in step $1$ does not introduce bias in large samples. Moreover, when the outcome model is correctly specified, this estimator is asymptotically efficient in the subclass of estimators that are unbiased as soon as randomisation is independent of baseline covariates $\bm{Z}$ \citep{Yang2001, Tsiatis2006, Moore2009, Stallard2010}.

The mean $E[Y^{l}|T=1, R=f]$, denoted as $\theta_2$, cannot be  estimated directly from the flexible dosing trial since $Y^l$ is not observed for patients in the flexible dosing arm of the flexible dosing trial. In view of this, we will transport data from  the low dose of the fixed dosing trial to the flexible dosing trial.
Sufficient conditions for identifying this mean by transporting inferences from the fixed dosing trial are mean exchangeability w.r.t. $T$, conditional on $\bm{Z}$ (transportability) and positivity of trial assignment. 
The first condition states that patients with the same characteristics $\bm{Z}$ in both trials should be comparable in terms of the outcome $Y^l$ that can be expected, i.e., $E(Y^l|T=1, \bm{Z})=E(Y^l|T=0, \bm{Z})=E(Y^l|\bm{Z})$.
%the counterfactual outcome under the low dose $Y^l$ has the same mean in both trials among patients with the same baseline covariates $\bm{Z}$, i.e.\,, $E(Y^l|T=1, \bm{Z})=E(Y^l|T=0, \bm{Z})=E(Y^l|\bm{Z})$.
The second condition states that in each stratum of those baseline covariates $\bm{Z}$, all patients have a positive probability to be assigned to each of the trials. This means that no patient, based on his/her characteristics $\bm{Z}$, is excluded from participating in each ones of the trials. Under these conditions and using that the flexible dosing trial is randomized, an estimator $\hat{\theta}_2$ for $\theta_2$ is obtained by \citep{Zhang2016, ShuTan2018, Dahabreh2018}
\begin{enumerate}
	\item fitting a parametric model $\pi(\bm{Z}, \bm{\gamma})$ for the probability of participation in the flexible dosing trial $P(T=1|\bm{X})$; e.g.\,, $\pi(\bm{X}, \bm{\gamma})=\text{logit}^{-1}(\gamma_0+\bm{\gamma}_1'\bm{Z})$, 
	\item fitting a weighted regression for the conditional mean of $Y$ on $\bm{Z}$ among the patients on the low dose of the fixed dosing trial using a canonical generalized linear working model with weights $\hat{\pi}(\bm{Z}_i, \hat{\bm{\gamma}})/(1-\hat{\pi}(\bm{Z}_i, \hat{\bm{\gamma}}))$: $E(Y|R=l, T=0, \bm{Z})=m(\bm{Z}, \bm{\beta}_0)$, where $m(\bm{Z}, \bm{\beta})$ is a known function, evaluated at a parameter $\bm{\beta}$ with unknown population value $\bm{\beta}_0$; e.g.\,, $m(\bm{Z}, \bm{\beta})=\beta_1+\bm{\beta}_2'\bm{Z}$ for a continuous outcome $Y$,
	\item taking the average of the predicted values over all patients in the flexible dosing trial.	
\end{enumerate}
This semi-parametric estimator is consistent when either the model for the selection $\pi(\bm{Z}, \bm{\gamma})$ or the outcome $m(\bm{Z}, \bm{\gamma})$ is correctly specified. Moreover, when both models are correctly specified, it achieves the non-parametric efficiency bound, under the non-parametric model where no assumptions are made about the outcome and selection model beyond the transportability and positivity assumptions \citep{Zhang2016, ShuTan2018, Dahabreh2018}.

It follows that Expression (\ref{eq:quest}) can be estimated as $(\hat{\theta}_1-\hat{\theta}_2)/\hat{\pi}^S$, with $\hat{\pi}^S$ the observed proportion of switchers in the flexible dosing arm. 
%This semi-parametric estimator has the appealing feature that only misspecification of both models used for $\hat{\theta_2}$, the model for the selection $\pi(\bm{Z}, \bm{\gamma})$ and the outcome model $m(\bm{Z}, \bm{\gamma})$, introduces bias.
This semi-parametric estimator has the appealing feature that misspecification of $h(\bm{Z}, \bm{\eta})$ as well as misspecification of at most one of the working models used for $\hat{\theta}_2$ (i.e.\,, $\pi(\bm{Z}, \bm{\gamma})$ and $m(\bm{Z}, \bm{\beta})$) does not introduce bias.
Moreover, it achieves the non-parametric efficiency bound when all models, $h(\bm{Z}, \bm{\eta})$, $\pi(\bm{Z}, \bm{\gamma})$ and $m(\bm{Z}, \bm{\gamma})$, are correctly specified. In order to calculate the variance of this estimator, one must take into account that the predictions used to calculate $\hat{\theta}_1$ and $\hat{\theta}_2$ are estimated based on particular outcome regression models.
It is therefore not sufficient to compute the sample variance of the mean of these predictions. In Appendix A.3, we show that this can be easily accommodated.

\subsubsection{Remark}\label{sec:remark}
Note that the component $E[Y^f-Y^{l}|T=1, R=f]$ of the treatment effect of interest (Equation (\ref{eq:quest})) is an average treatment effect on the ``treated'' (ATT) because it represents the mean difference of two potential outcomes over the subpopulation of individuals who were assigned to the flexible dose. 
In contrast with estimation of ATE, efficiency can be gained when the selection model is known since the selection model is not ancillary for estimation of (components of) ATT \citep{Hahn1998}.
%Since the selection model is not ancillary for estimation of (components of) ATT \citep{Hahn1998}, efficiency can be gained when the selection model is known, in contrast with estimation of ATE. 
A more efficient estimator for $\theta_1-\theta_2$ can in particular be obtained by \citep{ShuTan2018}
\begin{enumerate}
	\item fitting the same regression models $h(\bm{Z}, \bm{\eta})$ and $m(\bm{Z}, \bm{\beta})$, respectively, as before,
	\item taking the average of $\hat{\pi}(\bm{Z}_i, \hat{\gamma})\{\hat{h}(\bm{Z}_i; \hat{\eta})-\hat{m}(\bm{Z}_i; \hat{\beta})\}/\hat{\pi}^T$ over all patients,
	%\item taking the sum of respectively and $\hat{\pi}(\bm{Z}_i, \hat{\gamma})\hat{h}(\bm{Z}_i; \hat{\eta})$ and $\hat{\pi}(\bm{Z}_i, \hat{\gamma})\hat{m}(\bm{Z}_i; \hat{\beta})$ over all patients,
	%\item and dividing both by the number of patients in the flexible dosing trial.
\end{enumerate}
with $\hat{\pi}^T$ the estimated proportion of patients in the flexible dosing trial.
This results in a semi-parametric estimator for $\theta_1-\theta_2$ that achieves the semi-parametric efficiency bound,
under the semi-parametric model where the density of $T$ given baseline covariates $\bm{Z}$ is assumed to be known,
when the outcome models as well as the selection model are correctly specified \citep{ShuTan2018}. This variance bound calculated under the parametric selection model, is no greater than under the less restrictive model where the selection model is assumed to be unknown. Despite the fact that this estimator is more efficient, it is generally not unbiased when the selection model is misspecified and therefore no longer doubly robust.
Moreover, as shown in Appendix \ref{app_remark}, both estimators for $\theta$ are equivalent when using a logistic regression for $\pi(\bm{Z}, \bm{\gamma})$ with a set of covariates that includes the covariates used in the models $m(\bm{Z};\bm{\beta})$ and $h(\bm{Z};\bm{\eta})$.
 
In the motivating study, the selection model in unknown and we therefore recommend the doubly robust estimator. In Section \ref{sec:practcons} we discuss when it would be justified to use this more efficient estimator.

\section{Simulation Study}
We conducted a simulation study to compare the finite-sample performance of the proposed estimators versus a regression estimator \citep{ROBINS1986}. The latter estimator is obtained by estimating $\theta_1$ in the same way as before, and estimating $\theta_2$ by skipping step one, building an unweighted regression for $Y$ on $\bm{Z}$ for the patients in the low dose arm of the fixed dosing trial in step two and taking the average of the predicted values (based on the unweighted regression) over all patients in the flexible dosing trial in step three.

\subsection{Data Generation}
We considered $100$ patients in each arm, corresponding with $200$ patients in the flexible dosing trial and $300$ patients in the fixed dosing trial. 
We first generated (independent) baseline covariates in both trials as $X_{ji}\sim N(0, 1)$ for $j=1, 2, 3$ and $X_{ji}\sim Ber(0.5)$ for $j=4, 5, 6$; $i=1, \dots, n$. 
To allow for selection into the trials based on baseline covariates, we then generated two (independent) baseline covariates ($j=7, 8$) as $X_{ji}\sim N(0.5, 1)$ for patients in the flexible dosing trial and $X_{ji}\sim N(\mu, 1)$ for patients in the fixed dosing trial, and two (independent) baseline covariates ($j=9, 10$) as $X_{ji}\sim Ber(0.6)$ for patients in the flexible dosing trial and $X_{ji}\sim Ber(\phi)$ for patients in the fixed dosing trial. The difference in the
means of these covariate distributions of both trials, which is controlled by $\phi$ and $\mu$, represents selection into the trials based on baseline covariates.
To evaluate the strength of the selection, $5$ different combinations for ($\phi$, $\mu$) are considered; 
Setting $1$: ($0.6$, $0.5$), Setting $2$: ($0.5$, $0.25$), Setting $3$: ($0.4$, $0$), Setting $4$: ($0.2$, $-0.5$) and Setting $5$: ($0.1$, $-1$). 
The first setting representing no selection, and the last setting strong selection.

The potential outcomes ($Y^p$, $Y^l$, $Y^c$) are generated as respectively
$Y^p|\bm{X}\sim N(m_p(\bm{X}), 1)$, $Y^l|\bm{X}\sim N(m_l(\bm{X}), 1)$ and $Y^c|\bm{X}\sim N(m_c(\bm{X}), 1)$
with 
$$ m_p(\bm{X})= X_3+X_6+X_8+X_{10}+X_3X_6+X_7X_9,$$
$$ m_l(\bm{X})=X_1+X_2+ X_3+X_4+X_5+X_6+X_7+X_8+X_9+X_{10}+X_3X_6+X_7X_9,\text{ and}$$
$$ m_c(\bm{X})= -X_1-0.5X_2+ X_3-X_4-0.5X_5+X_6-0.5X_7+X_8-0.5X_9+X_{10}+X_3X_6+X_7X_9.$$
Moreover, $S$ was generated as $S|\bm{X}\sim Ber(\text{expit}(X_3+X_6+X_7+X_8+X_9+X_{10}))$.
This gives us, in all scenarios, a main treatment effect in the switchers of $-3.59$.

For estimation of the treatment effect, consider an outcome regression model with the identity link and regressor vectors
$(1, X_1, \dots, X_{10}, X_3X_6, X_7X_9)'$ and $(1, \log|X_1|, \log|X_2|,$ $\log|X_3|, X_4, X_5, X_6, \log|X_7|, \log|X_8|, X_9, X_{10})'$, corresponding to respectively a correctly specified and misspecfied outcome model. 
Similarly, consider a selection model with the logistic link and regressor vectors $(1, X_7, X_8, X_9, X_{10})'$ and $(1, \log|X_7|, \log|X_8|, X_9, X_{10})'$, corresponding to respectively a correctly specified and misspecfied selection model.
We estimated the performance of each estimator over $5000$ runs for each scenario.

\subsection{Simulation Results}
Table \ref{tab: results} summarizes the simulation results for the different settings under correctly specified working models, a correctly specified selection model but misspecified outcome models and correctly specified outcome models but a misspecified selection model. 
%	\centering		
%	\begin{threeparttable}
%		\caption{\label{tab: results}Comparison of G-computation and the estimators proposed in Section \ref{sec:proposal}.}
%		\begin{tabular}{ll l ccc}
%		\hline
%		Misspecification & Method & & Bias & SE & Coverage\\ \hline
%		Correctly specified & Non-parametric &&$-0.0002$&$0.0764$&$0.941$\\
%		& Semi-parametric &&$-0.0002$&$0.0761$&$0.939$\\
%		& G-computation &&$-0.0002$&$0.0751$&$0.944$\\
%		\hline
%		Outcome model misspecified & Non-parametric &&$0.0225$&$0.1002$&$0.943$\\
%		& Semi-parametric &&$0.0200$&$0.1031$&$0.943$\\
%		& G-computation &&$0.6430$&$0.1146$&$0.528$\\
%		\hline
%		Selection Model misspecified & Non-parametric &&$-0.0001$&$0.0757$&$0.942$\\
%		& Semi-parametric &&$-0.1614$&$0.0714$&$0.893$\\
%		& G-computation &&$-0.0002$&$0.0751$&$0.944$\\
%		\hline
%		\hline
%	\end{tabular}
%	\begin{tablenotes}
%	\small
%	\item Note: NP, eff. estimator under non-parametric model; SP, eff. estimator under under semi-parametric model. 
%\end{tablenotes}
%\end{threeparttable}
%\end{table}
\begin{landscape}
\begin{table}
	\centering		
	\begin{threeparttable}
		\caption{\label{tab: results}Comparison of the bias, the standard error and mean squared error of the estimator for the regression estimator and the estimators proposed in Section \ref{sec:proposal}.}
		\begin{tabular}{ll c c ccc c ccc c ccc}
			\hline
			 &  &  & & \multicolumn{3}{c}{Non-parametric} && \multicolumn{3}{c}{Semi-parametric} && \multicolumn{3}{c}{Regression}\\ \cline{5-7} \cline{9-11} \cline{13-15}
			Misspecification & Setting & Weights&&Bias&SE&MSE&&Bias&SE&MSE&&Bias&SE&MSE\\ \hline
			Correctly specified & Setting $1$ & $1.65; 2.34$ &&$-0.005$ & $0.110$ & $0.110$&&$-0.005$ &$0.087$ &$0.087$&&$-0.005$ &$0.110$ & $0.110$\\
			& Setting $2$ & $0.85; 3.33$ &&$-0.004$ & $0.114$ & $0.114$ &&$-0.003$& $0.093$&$0.093$&&$-0.003$& $0.114$ & $0.114$\\
			& Setting $3$ & $0.32; 4.01$ &&$-0.004$& $0.131$&$0.131$&&$-0.004$ &$0.112$&$0.112$&&$-0.004$ &$0.124$ & $0.124$\\
			& Setting $4$ & $0.03; 4.54$ &&$-0.007$& $0.317$&$0.317$&&$-0.007$ &$0.303$&$0.303$&&$-0.008$& $0.196$& $0.196$\\
			& Setting $5$ & $0.002; 2.53$ &&$-0.006$ & $1.473$&$1.473$&&$-0.005$ &$1.455$&$1.455$&&$-0.016$& $0.562$& $0.562$\\
			\hline
Outcome Model & Setting $1$ & $1.65; 2.34$ &&$0.013$ &$0.252$&$0.252$&&$0.008$& $0.236$&$0.236$&&$0.016$& $0.257$&$0.257$\\
misspecified& Setting $2$ & $0.85; 3.33$ &&$0.093$& $0.285$&$0.293$&&$0.084$& $0.267$&$0.274$&&$0.717$& $0.283$& $0.796$\\
& Setting $3$ & $0.32; 4.01$ &&$0.271$& $0.411$&$0.485$&&$0.258$ &$0.392$&$0.458$&&$1.526$& $0.329$&$2.658$\\
& Setting $4$ & $0.03; 4.54$ &&$1.318$& $1.190$&$2.926$&&$1.304$& $1.159$&$2.860$&&$3.049$& $0.558$&$9.854$\\
& Setting $5$ & $0.002; 2.53$ &&$2.965$& $2.735$&$11.523$&&$2.950$& $2.701$&$11.406$&&$4.060$ &$1.044$&$17.526$\\
\hline
Selection Model & Setting $1$ & $1.51; 3.06$ &&$-0.005$ &$0.110$&$0.110$&&$-0.004$& $0.084$&$0.084$&&$-0.005$ &$0.110$&$0.110$\\
misspecified& Setting $2$ & $1.23; 3.05$ &&$-0.004$ &$0.114$&$0.114$&&$0.265$& $0.090$&$0.161$&&$-0.003$ &$0.114$&$0.114$\\
& Setting $3$ & $0.77; 4.53$ &&$-0.003$ &$0.128$&$0.128$&&$0.534$& $0.108$&$0.394$&&$-0.004$ &$0.124$&$0.124$\\
& Setting $4$ & $0.46; 14.27$ &&$-0.010$ &$0.251$&$0.251$&&$0.850$& $0.212$&$0.934$&&$-0.008$& $0.196$&$0.196$\\
& Setting $5$ & $0.28; 6.41$ &&$-0.018$& $0.773$& $0.773$&&$0.897$& $0.613$&$1.418$&&$-0.016$& $0.562$ & $0.562$\\
\hline
			\hline
		\end{tabular}
\begin{tablenotes}
	\small
\item Note: The column weights shows the $5\%$ and $95\%$ percentiles of the weights $\hat{\pi}(\bm{Z}_i, \hat{\bm{\gamma}})/(1-\hat{\pi}(\bm{Z}_i, \hat{\bm{\gamma}}))$ among the patients on the low dose of the fixed dosing trial.
\end{tablenotes}
	\end{threeparttable}
\end{table}
\end{landscape}

When all models are correctly specified, all estimators are approximately unbiased. 
The efficient estimator under the non-parametric model has the highest variance. As long as the selection into both trials is not too strong (Settings $1$-$3$), the efficient estimator under the semi-parametric model has the lowest variance. This is a consequence of the fact that the predictions of all patients are used. 
Interestingly, the regression estimator is the most efficient in the presence of strong selection (Settings $4$ and $5$). 
This because strong selection and accordingly the extreme weights used in the proposed estimators increase the variance of these estimators.

When the outcome model is misspecified, the regression estimator is biased. The efficient estimators under non- and semi-parametric models are subject to small sample bias which disappears for larger samples (See Appendix Table \ref{tab: results_large}). Note that the bias becomes larger in the presence of stronger selection on covariates.
The variance of the two efficient estimators is comparable. 

When the selection model is misspecified, the efficient estimator under the semi-parametric model is biased since this approach assumes that the selection model is known and thus correctly specified. The variances are comparable between the different estimators. Note that the regression estimator is the same as under correctly specified models since it does not make use of the selection model.

Moreover, the semi-parametric estimator achieving the non-parametric efficiency bound seems to be most promising in practice. This because comparing the MSEs shows that the small efficiency gain of the other estimators does not exceed their sensitivity to bias.

Additional simulations were designed to evaluate the performance of the estimators when the mean exchangeability condition is violated (i.e., when there are unmeasured common causes of both the outcome and trial selection).
For estimation of the treatment effect, consider an outcome regression model with the identity link and regressor vector $(1, X_1, X_2, X_3, X_4, X_5, X_6, X_8, X_9, X_10, X_3X_6)'$ and a selection model with the logistic link and regressor vector $(1, X_8, X_9, X_{10})'$. Both models are missing the baseline covariate $X_7$ which is a common cause of outcome and trial selection. 
Moreover, we evaluate the performance of the estimators under violation of mean exchangeability but by assuming that another baseline covariate $X_{11}$, which is correlated with $X_7$, is measured.
We generated $X_{11}$ (and $X_7$) from a multivariate normal distribution with different covariances/correlations and mean and variance as described in the previous section.
For estimation of treatment effect, consider an outcome regression model with the identity link and regressor vector
$(1, X_1, \dots, X_6, X_8, \dots, X_{11}, X_3X_6, X_{11}X_9)'$ and a selection model with the logistic link and regressor vectors $(1, X_{11}, X_8, X_9, X_{10})'$.
Simulation results are presented in Appendix Tables \ref{tab: results_violated} and \ref{tab: results_violated2}.
When the mean exchangeability assumption is violated and there is selection into the trials, all the estimators are biased. This bias seems to be worse when the selection is stronger. However, when a baseline covariate $X_{11}$ is measured that is correlated with the omitted confounder $X_7$, the bias is less extreme, as expected.

\section{Illustration for Motivating Trials}
We analyse the data from a study program for a chronic condition at Janssen Pharmaceuticals that motivated this work.
%treatment-resistant depression at Janssen Pharmaceuticals that motivated this work.
For simplicity, we only report analyses restricted to patients in the full analysis set, on which the efficacy analyses were also performed.
%The full analysis set of the fixed dosing trial consists of $342$ patients; $115$ were assigned to the fixed low dose, $114$ to the fixed high dose and $113$ to (fixed) placebo. 
%One patient assigned to the fixed high dose had no complete data and was therefore excluded from the full analysis set used for the efficacy analyses,
%resulting in a dataset of $341$ patients.
%The full analysis set of the flexible dosing trial consists of $223$ patients; $114$ were assigned to the flexible dose, $109$ to (flexible) placebo. 
%Two patients assigned to the flexible dose had no complete data and were therefore excluded from the full analysis set used for the efficacy analyses, resulting in a dataset of $221$ patients.
In general, the treatment groups within trials were similar with respect to baseline and demographic characteristics. %(Figures \ref{fig:baseline_fixed} and \ref{fig:baseline_flex}).
The two trials differ in gender, 
%oral antidepressant 
treatment initiated prior to randomization (to which they had not responded) and demographic characteristics: 
the majority of subjects were female ($70.5\%$ in fixed dosing trial versus $61.9\%$ in flexible dosing trial) and were 
initiated with a treatment of category I
%oral antidepressant treatment with a serotonin and norepinephrine reuptake inhibitor 
($57.3\%$ in fixed dosing trial versus $68.2\%$ in flexible dosing trial), where category of treatment is divided into 2 types- Category I and II (ie, surgery and chemotherapy in oncology trials). The greatest percentage of subjects was white ($76.6\%$ in fixed dosing trial versus $93.3\%$ in flexible dosing trial) and 
not Hispanic or Latino ($62.5\%$ in fixed dosing trial versus $92.8\%$ in flexible dosing trial).
The subjects in the fixed dosing study were mainly enrolled in Europe ($24.9\%$) and North America ($45.3\%$),
while the subjects in the flexible dosing study were only enrolled in Europe ($60.1\%$) and North America ($39.9\%$).

We implemented the method described in Section \ref{sec:proposal} to estimate the treatment effect in the switchers of the felxible dosing trial, which
expresses how different the result would have been for them, had they been kept on the (fixed) low dose (i.e.\,, not switched).
The working model $h(\bm{Z}, \bm{\eta})$, obtained via forward selection at a $15\%$ significance level, 
for the outcome under the flexible dose ($A=f$) was a
linear model with covariates: country, weight at baseline, age at baseline and the baseline measurement of the outcome. 
Covariates were selected on a rather high significance level to guarantee selection of the (important) predictors of the outcome.
After performing backward elimination starting from a model with all three-way interactions at a $15\%$ significance level,
interactions were included for country and age, for weight and the baseline measurement of the outcome and for weight and age (See Appendix Table \ref{tab:regressionY2}).
This outcome model is then used for estimating $\theta_1$, as explained in Section \ref{sec:proposal}.
The working model for the outcome under the fixed low dose ($m(\bm{Z}, \bm{\beta})$) was obtained via forward selection followed by backward 
elimination starting from an (unweighted) model with all three-way interactions, both at a $15\%$ significance level. 
This resulted in an (unweighted) model with a three-way interaction between the baseline measurement of the outcome,
height at baseline and ethnicity and all lower order terms (See Appendix Table \ref{tab:regressionY}).
The working model for the probability of participation in the flexible dosing trial ($\pi(\bm{Z}, \bm{\gamma})$), was also obtained via 
forward selection followed by backward elimination at a $15\%$ significance level. 
Hereby, it is important to include outcome predictors that are also associated with participation to a certain trial,
in models for the outcome and probability in the flexible dosing trial \citep{Brookhart2006}. 
The selection was therefore conducted on the variables selected for the outcome model $m(\bm{Z}, \bm{\beta})$ only.
This resulted in a selection model only including ethnicity (See Appendix Table \ref{tab:regressionSelection}). 
Note that the categories ``unknown'' and ``unreported'' are merged to one category (``unknown'').
%
%An estimate for the treatment effect in the switchers is obtained by estimating $\theta_1$, $\theta_2$ and $P(S=1|T=1, R=f)$ as described in Section \ref{sec:proposal}. For estimating $\theta_1$, the model $h(\bm{Z}, \bm{\eta})$, as obtained by the selection procedure for the outcome under the flexible dose described above, is used.
For estimating $\theta_2$, a weighted regression for the outcome under the fixed low dose is fitted with weights based on the model $\pi(\bm{Z}, \bm{\gamma})$ that is obtained by the selection procedure for the selection model. 
The outcome model used in this weighted regression is the one obtained in the selection procedure for the outcome under the fixed low dose, where no weights were used.
%Note that in the selection procedure for the outcome model under the fixed low dose, no weights were used.
%Let denote the estimated values for $\bm{\eta}$ and $\bm{\gamma}$ using the models described above as $\hat{\bm{\eta}}$ and $\hat{\bm{\gamma}}$.
 %Then, an estimator for the treatment effect in the switchers is obtained by estimating $\theta_1$, using the model $h(\bm{Z}, \bm{\eta})$ as obtained by the selection procedure for the outcome under the flexible dose described above, and $\theta_2$, using a weighted model for $m(\bm{Z}, \bm{\beta})$ with weights $\pi(\bm{Z}, \hat{\bm{\gamma}})/(1-\pi(\bm{Z}, \hat{\bm{\gamma}}))$, as described in Section \ref{sec:proposal}. Note that in the selection procedure for the outcome model under the fixed low dose, no weights were used.
The obtained estimate $-4.311$ ($95\%$ CI: $[-9.404, 0.782]$; $p=0.096$) suggests that on average switching is beneficial compared to staying on the low dose for switchers of the flexible dosing trial.
%, albeit the difference is not significant. 
These results, which are directional but not significant, are likely influenced by inadequate sample size as the study was not powered for this (subgroup) analysis.
%It should be noted that this analysis is exploratory as the trial was not powered for this (subgroup) analysis and may, therefore, be underpowered.

When conducting this analysis, it was a concern that the transportability and positivity of trial assignment assumptions would not be met.
This was because both trials were mainly conducted in different countries, 
so that both trials differ in patient population with respect to demographic characteristics as country, race and ethnicity. 
Fortunately, these factors were not significantly associated with the outcome under the (fixed) low dose. 
As a result, there is some evidence that the difference between countries is relatively less important
and that positivity violation is not a major concern.
This may be because the adjustment for the baseline measurement of the outcome
%baseline depression score 
is already explaining the key differences between countries with respect to the clinical outcome.
Moreover, the estimated probabilities of particpation 
in the flexible dosing trial were far from one, meaning that there are no patients with a very 
low probability to be assigned to the fixed dosing trial. This is important to exclude near-positivity violations when transporting data from the fixed dosing trial (fixed low dose) to the flexible dosing trial. 
The distribution of these estimated probabilities was similar among patients in the flexible and fixed dosing trial, 
reflecting the fairly similar observed covariate distribution in both trials and the absence of
strong selection into the trial (at least based on ethnicity).

To investigate that the findings are not driven by model specification decisions,
it is valuable to conduct a sensitivity analysis.
We consider a sequence of different working models for the outcome model $m(\bm{Z}, \bm{\beta})$ and the selection model $\pi(\bm{Z}, \bm{\gamma})$, but only consider the model described in Appendix Table \ref{tab:regressionY2} for $h(\bm{Z}, \bm{\eta})$.
The selection model is extended by allowing the inclusion of variables that are not associated with the outcome. 
After forward selection at a $15\%$ significance level, a model including ethnicity, region, treatment initiated prior to randomization
%initiated oral antidepressant treatment 
and race was obtained. 
First, a model with interactions was considered. To prevent violation of the positivity assumption and in view of an inflated chance of false positives, forward selection to include interactions was performed at a $5\%$ significance level. This resulted in a selection model with a two-way interaction between ethnicity and race, and main effects for ethnicity, region, %initiated oral antidepressant treatment
treatment initiated prior to randomization and race. 
A second model did not include interactions because of the small amount of individuals in certain (combined) categories.
Besides the outcome model under the low fixed dose that was obtained by combining forward and backward selection (See Appendix Table \ref{tab:regressionY}), we also consider a working model for this outcome via forward selection at a $5\%$ significance level for the interactions. This resulted in a model only including the main effects for the baseline measurement of the outcome, height at baseline and ethnicity. Moreover, when using the working models for trial selection including variables that were only associated with trial selection (region, %initiated oral antidepressant treatment
treatment initiated prior to randomization and race), we also consider the (extended) working models for the outcome that also include the main effects of these variables.

A summary of the results is shown in Appendix Table \ref{tab: sensitivity} and Figure \ref{fig:ForestPlot}. 
%The estimates as well as the $p$-value in the sensitivity analysis seem to be more extreme than the one obtained in the main analysis. They however go in the same direction and the same decision at a one-sided $5\%$ significance level is made. This suggests that our findings are not driven by model specification decisions or violation of the assumptions. 
The obtained estimates are similar in terms of the direction of the effect, but the sensitivity analysis seems to reinforce the obtained results of the primary analysis.
Moreover, given the imprecision of the estimates, the choice of the models does not impact the size of the effect and the corresponding confidence intervals too strongly. 
The imprecision of the estimates is likely due to lack of overlap in the data with respect to demographic characteristics as country/region, race and ethnicity.

\section{Practical Considerations for Trial Designs}
\label{sec:practcons}
The considered problem raises the question whether the design should be adapted to the question of interest. For example, a trial with five arms where patients are randomized to either the low dose, high dose, flexible dosing, fixed placebo or flexible placebo, would result in a simpler estimator since there would be no need to transport between two studies. Here, we suggest to use two placebo arms, a flexible and a fixed placebo arm, to make it better possible to estimate the effect of flexible dosing compared to fixed dosing. Note that the dose in such a trial would be blinded, but not whether a patient is assigned to a fixed or a flexible dosing not. This type of design, however, would contradict the demand of the authorities for two positive efficacy trials. 
Another preferable possibility may therefore be to conduct two separate trials, but making them comparable by randomizing patients, in a simple or stratified way, between the two trials. By randomization, the selection model is then known and the more efficient estimator proposed in Section \ref{sec:remark} can be used. However, the feasibility of such a trial need further study.

When using a similar design as the motivating example, which seems most realistic in practice, it is important to ensure that the patients are properly selected to prevent structural violations of the positivity of trial assignment assumption. This is satisfied when the observed covariate distributions in the fixed and flexible dosing trial are fairly similar.
This can, for example, be done by matching by design; e.g.\,, for every patient in the flexible dosing trial, finding a patient with similar observable characteristics to be assigned to the fixed dosing trial, or vice versa. 
It may also be that it is not feasible to conduct both studies in the same countries, resulting in a similar situation of non-overlapping countries as in the trials which motivated this work.
It can then be helpful to randomize countries itself over the two trials. Meaning that all people from the same country participate in the same trial, so that one still conduct two separate studies in two contexts, but one can nevertheless expect a certain comparability.

Independent of the design, it is always a good idea to examine whether the distribution of the estimated probabilities of being selected for the flexible dosing trial (or fixed dosing trial) are similar among patients in the flexible dosing trial and patients in the fixed dosing trial. This would reflect the fairly similar observed covariate distributions in the fixed dosing trial and flexible dosing trial as well as the absence of strong selection into one of the trials, at least based on the measured covariates.

\section{Discussion}
\label{sec:disc}
In this paper, we have proposed a method to estimate the treatment effect in the switchers, which
expresses how different the result would have been for them, had they been kept on the low dose.
To realize this, data from a fixed dosing trial that has been conducted concurrently
on the same target, albeit not in an identical patient population, was transported.
We formulated the statistical problem in terms of potential outcomes and discuss the plausibility
of the assumptions required to transport the mean of the potential outcomes under
a given treatment regimen from one trial to another. To avoid undue reliance on
modelling assumptions, we proposed a doubly robust estimator, which relies on an
outcome model and a propensity score model for the association between trial and
patient characteristics but only requires one of them to be correcly specified. 

A major challenge in applying the proposed method (and methods to transport data in general) is the need
to collect adequate information on covariates for the mean transportability assumption to hold. 
Although the transportability assumption is untestable, Pearl and Bareinboim have recently proposed methods to facilitate assessing its plausibility \citep{Bareinboim2012, Bareinboim2013, Pearl2014}.
Moreover, it is important to ensure that the patients are properly selected to prevent 
structural violations of the positivity of trial assignment assumption.
For example, if one of the trials restricted enrollment to patients under $70$ years of age, it is prudent to apply the same
restriction in the other trial.

The fact that the trials in the motivating example were not conducted in the same countries and/or centers, raised the question whether the transportability and positivity assumptions were met. 
This because of investigators in different countries/centers may handle some situations (e.g.\,, switching to a different dose) differently, and the possible differences in demographic characteristics (e.g.\,, country, race, ethnicity, etc.) and culture may play an important role in the evaluation of certain outcomes. 
%This because of investigators in different countries/centers may handle some situations (e.g.\,, switching to a different dose) differently, culture may play an important role in an outcome for depression compared in contrast to an outcome such as blood pressure and the possible differences in demographic characteristics (e.g.\,, country, race, ethnicity, etc.). 
However, it seemed that the adjustment for %baseline depression score
the baseline measurement of the outcome was already explaining the key differences between countries with respect to the clinical outcome. Moreover, similar transportability assumptions are made implicitly in meta-analyses, where they are also considered reasonable, even if the studies were conducted in different studies.
%Similarly, both trials should be conducted in the same countries, and even centers.
%This because of investigators in different countries/centers may handle some situations (e.g.\,, switching to a different dose) differently, culture may play an important role in an outcome for depression compared in contrast to an outcome such as blood pressure
%and the possible differences in demographic characteristics (e.g.\,, country, race, ethnicity, etc.).
%Since this was not the 
%case for the trials used for the data analysis, transportability and positivity of trial assignment assumptions may not be met.
Nevertheless, in practice, one should select the patients in such a way 
(e.g.\,, by matching or randomization over the trials) that the observed covariate distributions are fairly similar 
and that there is no strong selection into one of the trials in order to guarantee valid inferences based on the proposed method.

Simulation results have shown that it is more appropriate to use the proposed method using doubly robust estimators 
than using the regression estimator since the double robustness property gives us two opportunities for correct inference.
In adition, the semi-parametric estimator achieving the non-parametric efficiency bound is more promising in practice
than the one achieving the semi-parametric efficiency bound since the latter is generally not unbiased 
when the selection model is misspecified and moreover the efficiency gain is minor.  

Further research is also needed to examine the behavior of the proposed estimators under model misspecication, as well as to improve their efficiency and stability under misspecifiaction of the outcome or selection model \citep{Robins2007, Cao2009, Vermeulen2015}.

\newpage
%\bigskip
\begin{center}
{\large\bf SUPPLEMENTARY MATERIAL}
\end{center}
\section*{Appendix A: proposed estimator}
The focus in this section is on an estimator for $E[Y^{c}-Y^{l}|T=1, R=f, S=1]$, or equivalently $E[Y^f-Y^{l}|T=1, R=f]/P(S=1|T=1, R=f)$, and its properties. 

\subsection*{Appendix A.1: Theoretical Derivation of the Estimator}
Let $n_1$ and $n_0$, with $n=n_1+n_0$, be the sample sizes in respectively the flexible and fixed dosing trial. Moreover, $n_1=\pi^Tn$, with $\pi^T=P(T=1)$. 

First, following the reasoning in Tsiatis \cite{Tsiatis2006}, it follows that $\theta_1= E[Y^{f}|T=1, R=f]$ can be estimated as
\begin{align*}
\hat{\theta}_{1} 
%&=\frac{\sum_{i=1}^{n} \frac{T_i}{\hat{\pi}_T}\left\{ \frac{I(R_i=f)}{\hat{\pi}^f}Y_i+\left(1-\frac{I(R_i=f)}{\hat{\pi}^f}\right)\hat{Y}_{fi}\right\}}{\sum_{i=1}^{n} \frac{T_i}{\hat{\pi}_T}}\\
%&=n_1^{-1}\sum_{i=1}^{n_1} \frac{I(R_i=f)}{\hat{\pi}^f}Y_i+\left(1-\frac{I(R_i=f)}{\hat{\pi}^f}\right)\hat{Y}_{fi}\\
&=n_1^{-1}\sum_{i=1}^{n_1}  \frac{I(R_i=f)}{\hat{\pi}^f}\left(Y_i-h(\bm{Z}_i, \bm{\hat{\eta}})\right)+h(\bm{Z}_i, \bm{\hat{\eta}}).
\end{align*}
By estimating $\hat{\bm{\eta}}$ as the solution to the equations $\sum_{i=1}^n\frac{\text{I}(A_i=f)}{\hat{\pi}^f}(1, \bm{Z}_i)'\left(Y_i-h(\bm{Z}_i;\bm{\eta})\right)=0$, it follows that the estimator $\hat{\theta}_{1}$ reduces to 
$n_1^{-1}\sum_{i=1}^{n_1}h(\bm{Z}_i, \bm{\hat{\eta}})$; which coincides with the estimator proposed for $\theta_1$ in the main paper.

Next, using the reasoning in \cite{Zhang2016} and \cite{Dahabreh2018}, it follows that $\theta_2= E[Y^{l}|T=1, R=f]$ can be estimated as
\begin{align*}
\begin{split}
\hat{\theta}_{2}=&\sum_{i=1}^n\left[\frac{\text{I}(T_i=0)\text{I}(R_i=l)}{(1-\pi(\bm{Z}_i,\hat{\bm{\gamma}}))\hat{\pi}^l}\pi(\bm{Z}_i,\hat{\bm{\gamma}})\left(Y_i-m(\bm{Z}_i;\hat{\bm{\beta}})\right)+T_im(\bm{Z}_i;\hat{\bm{\beta}})\vphantom{\frac{P(S=1|X)}{P(S=0|X)}} \right]\Biggm/ \sum_{i=1}^nT_i.
\end{split}
\end{align*}
As proven in \cite{Zhang2016} this estimator achieves the non-parametric efficiency bound.  
By estimating $\hat{\bm{\beta}}$ as the solution to the equations\\ $\sum_{i=1}^n\frac{\text{I}(T_i=0)\text{I}(R_i=l)}{(1-\pi(\bm{Z}_i,\hat{\bm{\gamma}}))\hat{\pi}^l}\pi(\bm{Z}_i,\hat{\bm{\gamma}})(1, \bm{Z}_i)'\left(Y_i-m(\bm{Z}_i;\bm{\beta})\right)=0$, it follows that the estimator $\hat{\theta}_{2}$ reduces to 
$\sum_{i=1}^n\left(T_im(\bm{Z}_i;\hat{\bm{\beta}})\right)/\sum_{i=1}^nT_i$; which coincides with the estimator proposed for $\theta_2$ in the main paper. 

\subsection*{Appendix A.2: (Double) Robustness of the Estimator}
Because of simple randomisation, $\hat{\pi}^f$ and $\hat{\pi}^l$ are consistent estimator for $P(R=f|T=1)$ and $P(R=l|T=0)$. Additionally, $\hat{\pi}^T$ is a consistent estimator for $P(T=1)$. Define the probability limits $\bm{\eta^*}=\text{plim}(\bm{\hat{\eta}})$, $\bm{\beta^*}=\text{plim}(\bm{\hat{\beta}})$ and $\bm{\gamma^*}=\text{plim}(\bm{\hat{\gamma}})$, which equal respectively the true values $\bm{\eta}_0$,  $\bm{\beta}_0$ and $\bm{\gamma}_0$ when the working models $h(\bm{Z}, \bm{\eta})$, $m(\bm{Z}, \bm{\beta})$ and $\pi(\bm{Z}, \bm{\gamma})$ are correctly specified, but not necessarily otherwise. 

First, by the weak law of large numbers
and Slutsky's theorem, $\hat{\theta}_1$ estimates 
\begin{align*}
E&\left[\frac{TI(R=f)}{\pi^T\pi^f}Y+\frac{T}{\pi^T}\left(1-\frac{I(R=f)}{\pi^f}\right)h(\bm{Z}, \bm{\eta^*})\right]\\
=E&\left[E\left\{\frac{TI(R=f)}{\pi^T\pi^f}Y+\frac{T}{\pi^T}\left(1-\frac{I(R=f)}{\pi^f}\right)h(\bm{Z}, \bm{\eta^*})\middle|T=1, R=f,\bm{Z}\right\}\middle|T=1, R=f\right]\\
=E&\left[E\left\{Y\middle|T=1, R=f,\bm{Z}\right\}\middle|T=1, R=f\right]\\
=E&\left[Y\middle|T=1, R=f\right].
\end{align*}
We therefore describe the proposed interim estimator as robust, in the sense that it is consistent even if the outcome model $h(\bm{Z}, \bm{\eta})$ for $E(Y|R=f, T=1, \bm{Z})$ is misspecified (Tsiatis, $2008$).

Next, we show that $\hat{\theta}_2$ has the appealing future that misspecification of at most one of the two working models does no introduce bias. 
First, consider the case where $\pi(\bm{Z};\bm{\gamma})$ is correctly specified and $m(\bm{Z};\bm{\beta})$ (possibly) incorrectly specified:
\begin{align*}
&E\left[\frac{I(T=0)I(R=l)}{\left\{1-\pi(\bm{Z};\bm{\gamma}_0)\right\}\pi^l}\pi(\bm{Z};\bm{\gamma}_0)\left\{Y-m(\bm{Z};\bm{\beta^*})\right\}+T\cdot m(\bm{Z};\bm{\beta^*})\right]\Biggm/ E\left\{T\right\}\\
&=E\left[\left\{T-\frac{I(T=0)I(R=l)}{\left\{1-\pi(\bm{Z};\bm{\gamma}_0)\right\}\pi^l}\pi(\bm{Z};\bm{\gamma}_0)\right\}m(\bm{Z};\bm{\beta^*})+\frac{I(T=0)I(R=l)}{\left\{1-\pi(\bm{Z};\bm{\gamma}_0)\right\}\pi^l}\pi(\bm{Z};\bm{\gamma}_0)Y\right]\Biggm/ E\left\{T\right\}\\
&=E\left[\frac{I(T=0)I(R=l)}{\left\{1-\pi(\bm{Z};\bm{\gamma}_0)\right\}\pi^l}\pi(\bm{Z};\bm{\gamma}_0)Y\right]\Biggm/ E\left\{T\right\}\\
&=E\left[\pi(\bm{Z};\bm{\gamma}_0) E\{Y|T=0, R=l, \bm{Z}\}\right]/E\left\{T\right\}\\
&=E\left[\pi(\bm{Z};\bm{\gamma}_0) E\{Y^l|T=1, \bm{Z}\}\right]/E\left\{T\right\}\\
&=E\left[E\{TY^l|\bm{Z}\}\right]/E\left\{T\right\}\\
&=E\left[TY^l\right]/E\left\{T\right\}\\
&=E\left[ Y^{l}|T=1\right].
\end{align*}
Second, consider the case where $m(\bm{Z};\bm{\beta})$ is correctly specified and $\pi(\bm{X};\bm{\gamma})$ (possibly) incorrectly specified:
\begin{align*}
&E\left[\frac{I(T=0)I(R=l)}{\left\{1-\pi(\bm{Z};\bm{\gamma^*})\right\}\pi^l}\pi(\bm{Z};\bm{\gamma^*})\left\{Y-m(\bm{Z};\bm{\beta}_0)\right\}+T\cdot m(\bm{Z};\bm{\beta}_0)\right]\Biggm/ E\left\{T\right\}\\
&=E\left[T\cdot m(\bm{Z};\bm{\beta_0})\right]/E\left\{T\right\}\\
&=E\left[ Y^{l}|T=1\right].
\end{align*}
Thus, this doubly robust estimator is consistent when either the model for the propensity score or the expectation of the outcome is correctly specified.

Concludingly, the proposed semi-parametric estimator has the appealing feature that misspecification of $h(\bm{Z}, \bm{\eta})$ as well as misspecification of at most one of the working models used for $\hat{\theta}_2$ (i.e.\,, $\pi(\bm{Z}, \bm{\gamma})$ and $m(\bm{Z}, \bm{\beta})$) does not introduce bias. Thus, the proposed estimator will only be biased if both models used for $\hat{\theta_2}$, the model for the selection $\pi(\bm{Z}, \bm{\gamma})$ and the outcome model $m(\bm{Z}, \bm{\gamma})$, are misspecified.

\subsection*{Appendix A.3: Asymptotic Variance of the Estimator}\label{App:var}
Since the asymptotic behavior of an estimator is fully determined by
its influence function, it suffices to focus on
the influence function when discussing the estimator's variance. Denote the parameter of interest, $E[Y^{c}-Y^{l}|T=1, R=f, S=1]$, as $\theta$ and its estimator based on $\hat{\theta}_1$ and $\hat{\theta_2}$ as $\hat{\theta}$. 
%Note that $\theta=(\theta_1-\theta_2)/\pi^S$, with $\pi^S=P(S=1|T=1, R=f)$.
Define 
\begin{align*}
\begin{split}
\phi(Y, \bm{Z}, T, R; \theta, \bm{\eta}, \bm{\gamma}, \bm{\beta}, \pi^f, \pi^l, \pi^T, \pi^S)&=\frac{1}{\pi^T}\left[T\left\{ \frac{I(R=f)}{\pi^f}(Y-h(\bm{Z}, \bm{\eta}))+h(\bm{Z}, \bm{\eta})\right\}\right.\\
&-\left\{\frac{I(T=0)I(R=l)}{\pi^l}\exp(\bm{\gamma}^{T}\bm{Z})(Y-m(\bm{Z};\bm{\beta}))\right.\\
&+\left.\left.Tm(\bm{Z};\bm{\beta})\vphantom{\frac{A}{B}}\right\}\right]\Biggm/\pi^S-\frac{T}{\pi^T}\theta.
\end{split}
\end{align*}
Under the assumption that $\hat{\bm{\eta}}$, $\hat{\bm{\gamma}}$, $\hat{\bm{\beta}}$, $\hat{\pi}^f$ $\hat{\pi}^l$, $\hat{\pi}^T$ and $\hat{\pi}^S$ are the solutions to respectively the estimating equations  ${n}^{-1}\sum_{i=1}^{n}\bm{U_\eta}(Y_i, \bm{Z_i}, T_i, R_i; \bm{\eta})=\bm{0}$ with $\bm{U_\eta}(Y, \bm{Z}, T, R; \bm{\eta})=\bm{Z}(Y-\bm{\eta^T Z})TI(R=f)/(\pi^T\pi^{f})$, 
${n}^{-1}\sum_{i=1}^{n}\bm{U_\gamma}(\bm{Z}_i, T_i; \bm{\gamma})=\bm{0}$ with $\bm{U_\gamma}(\bm{Z}, T; \bm{\gamma})=\bm{Z}\{T-\text{expit}(\bm{\gamma^T Z})\}$,
${n}^{-1}\sum_{i=1}^{n}\bm{U_\beta}(Y_i, \bm{Z}_i, T_i, R_i; \bm{\beta})=\bm{0}$ with $\bm{U_\beta}(Y, \bm{Z}, T, R; \bm{\beta})=\bm{Z}(Y-\bm{\beta^T Z})I(T=0)I(R=l)\exp(\bm{\gamma}^T \bm{Z})/(\pi^{l}\pi^{T})$,
${n}^{-1}\sum_{i=1}^{n}\bm{U_{\pi^f}}(T_i, R_i; \pi^{f})=0$ with $\bm{U_{\pi^f}}(T, R; \pi^f)=T(I(R=f)-\pi^f)$,
${n}^{-1}\sum_{i=1}^{n}\bm{U_{\pi^l}}(T_i, R_i; \pi^{l})=0$ with $\bm{U_{\pi^l}}(T, R; \pi^l)=I(T=0)(I(R=l)-\pi^l)$,
 ${n}^{-1}\sum_{i=1}^{n}\bm{U_{\pi^T}}(T_i; \pi^T)=0$ with $\bm{U_{\pi^T}}(T;\pi^T)=T-\pi^T$ and
${n}^{-1}\sum_{i=1}^{n}\bm{U_{\pi^S}}(T_i, R_i, S_i; \pi^{S})=0$ with $\bm{U_{\pi^S}}(T, R, S; \pi^S)=TI(R=f)(S-\pi^S)$, 
the influence function of $\hat{\theta}$ (Tsiatis, $2006$) is given by 
\begin{align*}
\tilde{\phi}(Y, &\bm{Z}, T, R; \theta, \bm{\eta}^*, \bm{\gamma}^*, \bm{\beta}^*, \pi^f, \pi^l, \pi^T, \pi^S) =\phi(Y, \bm{Z}, T, R; \theta, \bm{\eta}^*, \bm{\gamma}^*, \bm{\beta}^*, \pi^f, \pi^l, \pi^T, \pi^S)\\
-&E\left\{\frac{\partial \phi}{\partial \bm{\eta}^T}(Y, \bm{Z}, T, R; \theta, \bm{\eta}^*, \bm{\gamma}^*, \bm{\beta}^*, \pi^f, \pi^l, \pi^T, \pi^S) \right\}E^{-1}\left\{\frac{\partial \bm{U_\eta}}{\partial \bm{\eta}^T}(Y, \bm{Z}, T, R; \bm{\eta^*})\right\}\bm{U_\eta}(Y, \bm{Z}, T, R; \bm{\eta^*})\\
-&E\left\{\frac{\partial \phi}{\partial \bm{\gamma}^T}(Y, \bm{Z}, T, R; \theta, \bm{\eta}^*, \bm{\gamma}^*, \bm{\beta}^*, \pi^f, \pi^l, \pi^T, \pi^S) \right\}E^{-1}\left\{\frac{\partial \bm{U_\gamma}}{\partial \bm{\gamma}^T}(\bm{Z}, T; \bm{\gamma^*})\right\}\bm{U_\gamma}(\bm{Z}, T; \bm{\gamma^*})\\
-&E\left\{\frac{\partial \phi}{\partial \bm{\beta}^T}(Y, \bm{Z}, T, R; \theta, \bm{\eta}^*, \bm{\gamma}^*, \bm{\beta}^*, \pi^f, \pi^l, \pi^T, \pi^S) \right\}E^{-1}\left\{\frac{\partial \bm{U_\beta}}{\partial \bm{\beta}^T}(Y, \bm{Z}, T, R; \bm{\beta^*})\right\}\bm{U_\beta}(Y, \bm{Z}, T, R; \bm{\beta^*})\\
-&E\left\{\frac{\partial \phi}{\partial \pi^f}(Y, \bm{Z}, T, R; \theta, \bm{\eta}^*, \bm{\gamma}^*, \bm{\beta}^*, \pi^f, \pi^l, \pi^T, \pi^S) \right\}E^{-1}\left\{\frac{\partial \bm{U_{\pi^f}}}{\partial \pi^f}(T, R; \pi^f)\right\}\bm{U_{\pi^f}}(T, R; \pi^f)\\
-&E\left\{\frac{\partial \phi}{\partial \pi^l}(Y, \bm{Z}, T, R; \theta, \bm{\eta}^*, \bm{\gamma}^*, \bm{\beta}^*, \pi^f, \pi^l, \pi^T, \pi^S) \right\}E^{-1}\left\{\frac{\partial \bm{U_{\pi^l}}}{\partial \pi^l}(T, R; \pi^l)\right\}\bm{U_{\pi^l}}(T, R; \pi^l)\\
-&E\left\{\frac{\partial \phi}{\partial \pi^T}(Y, \bm{Z}, T, R; \theta, \bm{\eta}^*, \bm{\gamma}^*, \bm{\beta}^*, \pi^f, \pi^l, \pi^T, \pi^S) \right\}E^{-1}\left\{\frac{\partial \bm{U_{\pi^T}}}{\partial \pi^T}(T; \pi^T)\right\}\bm{U_{\pi^T}}(T; \pi^T)\\
-&E\left\{\frac{\partial \phi}{\partial \pi^S}(Y, \bm{Z}, T, R; \theta, \bm{\eta}^*, \bm{\gamma}^*, \bm{\beta}^*, \pi^f, \pi^l, \pi^T, \pi^S) \right\}E^{-1}\left\{\frac{\partial \bm{U_{\pi^S}}}{\partial \pi^S}(T, R, S; \pi^S)\right\}\bm{U_{\pi^S}}(T, R, S; \pi^S)\\.
\end{align*}
First,
\begin{align*}
E\left\{\frac{\partial \phi}{\partial \bm{\eta}^T}(Y, \bm{Z}, T, R; \theta, \bm{\eta}^*, \bm{\gamma}^*, \bm{\beta}^*, \pi^f, \pi^l, \pi^T, \pi^S)\right\}=E\left\{\frac{1}{\pi^S}\frac{T}{\pi^T}
\left(1-\frac{I(R=f)}{\pi^f}\right)h'(\bm{Z}, \bm{\eta}^*)\right\}=\bm{0}^T,
\end{align*}
by the fact that $\pi^f$ is guaranteed to be correctly specified under simple randomisation. 
Next,
\begin{align*}
E\left\{
\frac{\partial \phi}{\partial \bm{\gamma}^T}(Y, \bm{Z}, T, R; \theta, \bm{\eta}^*, \bm{\gamma}^*, \bm{\beta}^*, \pi^f, \pi^l, \pi^T, \pi^S)\right\}&=-E\left\{\frac{1}{\pi^S}
\frac{I(T=0)I(R=l)}{\pi^T\pi^l}\exp(\bm{\gamma}^{*T}\bm{Z})\right.\\
&\left.\left\{Y-m(\bm{Z};\bm{\beta}^*)\right\}\bm{Z}^T\vphantom{\frac{A}{B}}\right\}=\bm{0}^T,
\end{align*}
and
\begin{align*}
E\left\{\frac{\partial \phi}{\partial \pi^l}(Y, \bm{Z}, T, R; \theta, \bm{\eta}^*, \bm{\gamma}^*, \bm{\beta}^*, \pi^f, \pi^l, \pi^T, \pi^S) \right\}&=E\left\{\frac{1}{\pi^S}
\frac{I(T=0)I(R=l)}{\pi^T(\pi^l)^2}\exp(\bm{\gamma}^{*T}\bm{Z})\right.\\
&\left.\left\{Y-m(\bm{Z};\bm{\beta}^*)\right\}\bm{Z}^T\vphantom{\frac{A}{B}}\right\}\\&=0,
\end{align*}
because of the way $\bm{\beta}$ is estimated. Next,
\begin{align*}
E\left\{\frac{\partial \phi}{\partial \pi^f}(Y, \bm{Z}, T, R; \theta, \bm{\eta}^*, \bm{\gamma}^*, \bm{\beta}^*, \pi^f, \pi^l, \pi^T, \pi^S) \right\}&=-E\left\{\frac{1}{\pi^S}\frac{T}{\pi^T} \frac{I(R=f)}{(\pi^f)^2}(Y-h(\bm{Z}, \bm{\eta}^*))\right\}=0,
\end{align*}
because of the way $\bm{\eta}$ is estimated.
Next,
\begin{align*}
\begin{split}
E\left\{\frac{\partial \phi}{\partial \pi^T}(Y, \bm{Z}, T, R; \theta, \bm{\eta}, \bm{\gamma}, \bm{\beta}, \pi^f, \pi^l, \pi^T, \pi^S)\right\}&=-E\left[\frac{1}{\pi^S}\frac{1}{(\pi^T)^2}\left\{T\left( \frac{I(R=f)}{\pi^f}(Y-h(\bm{Z}, \bm{\eta}))+h(\bm{Z}, \bm{\eta})\right)\right.\right.\\
&-\left(\frac{I(T=0)I(R=l)}{\pi^l}\exp(\bm{\gamma}^{T}\bm{Z})(Y-m(\bm{Z};\bm{\beta}))\right.\\
&+\left.\left.\left.Tm(\bm{Z};\bm{\beta})\vphantom{\frac{A}{B}}\right)\right\}-\frac{T}{(\pi^T)^2}\theta\right]\\
&=-E\left[\frac{1}{\pi^S}\frac{T}{(\pi^T)^2}\left\{h(\bm{Z}, \bm{\eta})-m(\bm{Z};\bm{\beta})\right\}-\frac{T}{(\pi^T)^2}\theta\right]\\
&=0,
\end{split}
\end{align*}
because of the way $\theta$ is estimated.
Finally, 
\begin{align*}
E\left\{\frac{\partial \phi}{\partial \bm{\beta}^T}(Y, \bm{Z}, T, R; \theta, \bm{\eta}, \bm{\gamma}, \bm{\beta}, \pi^f, \pi^l, \pi^T, \pi^S)\right\}&=-E\left\{
\frac{1}{\pi^S}\frac{1}{\pi^T}\left(\vphantom{\frac{A}{B}}I(T=1)\right.\right.\\
&\left.\left.-I(T=0)\frac{I(R=l)}{\pi^l}\exp(\bm{\gamma}^{*T}\bm{Z})\right)\bm{Z}^T\right\}
\end{align*}
and
\begin{align*}
E\left\{\frac{\partial \phi}{\partial \pi^S}(Y, \bm{Z}, T, R; \theta, \bm{\eta}, \bm{\gamma}, \bm{\beta}, \pi^f, \pi^l, \pi^T, \pi^S)\right\}&=-\frac{1}{(\pi^S)^2}
\frac{1}{\pi^T}E\left\{T\left( \frac{I(R=f)}{\pi^f}(Y-h(\bm{Z}, \bm{\eta}))+h(\bm{Z}, \bm{\eta})\right)\right.\\
&-\left(\frac{I(T=0)I(R=l)}{\pi^l}\exp(\bm{\gamma}^{T}\bm{Z})(Y-m(\bm{Z};\bm{\beta}))\right.\\
&+\left.\left.Tm(\bm{Z};\bm{\beta})\vphantom{\frac{A}{B}}\right)\right\}.
\end{align*}
Therefore, the influence function of the parameter of interest, $\theta$, reduces to
\begin{align*}
\phi(Y,& \bm{Z}, T, R; \theta, \bm{\eta}, \bm{\gamma}, \bm{\beta}, \pi^f, \pi^l, \pi^T, \pi^S)\\
-&E\left\{\frac{\partial \phi}{\partial \bm{\beta}^T}(Y, \bm{Z}, T, R; \theta, \bm{\eta}, \bm{\gamma}, \bm{\beta}, \pi^f, \pi^l, \pi^T, \pi^S) \right\}E^{-1}\left\{\frac{\partial \bm{U_\beta}}{\partial \bm{\beta}^T}(Y, \bm{Z}, T, R; \bm{\beta})\right\}\bm{U_\beta}(Y, \bm{Z}, T, R; \bm{\beta})\\
-&E\left\{\frac{\partial \phi}{\partial \pi^S}(Y, \bm{Z}, T, R; \theta, \bm{\eta}, \bm{\gamma}, \bm{\beta}, \pi^f, \pi^l, \pi^T, \pi^S) \right\}E^{-1}\left\{\frac{\partial \bm{U_{\pi^S}}}{\partial \pi^S}(T, R, S; \pi^S)\right\}\bm{U_{\pi^S}}(T, R, S; \pi^S),
\end{align*}
with $E^{-1}\left\{\frac{\partial \bm{U_\beta}}{\partial \bm{\beta}^T}(Y, \bm{Z}, T, R; \bm{\beta})\right\}=E^{-1}\left\{-I(T=0)I(R=l)\exp(\bm{\gamma}^T \bm{Z})/(\pi^{l}\pi^{T})\bm{Z}\bm{Z}^T\right\}$ and  
$E^{-1}\left\{\frac{\partial \bm{U_{\pi^S}}}{\partial \pi^S}(T, R, S; \pi^S)\right\}=-E^{-1}\{TI(R=f)\}$.
The asymptotic variance of $\hat{\theta}=(\hat{\theta}_1-\hat{\theta}_{2})/\hat{\pi}^S$ can then be easily calculated as one over $n$ times the sample variance of
\begin{align*}
\phi(Y,& \bm{Z}, T, R; \hat{\theta}, \hat{\bm{\eta}}, \hat{\bm{\gamma}}, \hat{\bm{\beta}}, \hat{\pi}^f, \hat{\pi}^l, \hat{\pi}^T, \hat{\pi}^S)\\
-&E\left\{\frac{\partial \phi}{\partial \bm{\beta}^T}(Y, \bm{Z}, T, R; \hat{\theta}, \hat{\bm{\eta}}, \hat{\bm{\gamma}}, \hat{\bm{\beta}}, \hat{\pi}^f, \hat{\pi}^l, \hat{\pi}^T, \hat{\pi}^S) \right\}E^{-1}\left\{\frac{\partial \bm{U_\beta}}{\partial \bm{\beta}^T}(Y, \bm{Z}, T, R; \hat{\bm{\beta}})\right\}\bm{U_\beta}(Y, \bm{Z}, T, R; \hat{\bm{\beta}})\\
-&E\left\{\frac{\partial \phi}{\partial \pi^S}(Y, \bm{Z}, T, R; \hat{\theta}, \hat{\bm{\eta}}, \hat{\bm{\gamma}}, \hat{\bm{\beta}}, \hat{\pi}^f, \hat{\pi}^l, \hat{\pi}^T, \hat{\pi}^S) \right\}E^{-1}\left\{\frac{\partial \bm{U_{\pi^S}}}{\partial \pi^S}(T, R, S; \hat{\pi}^S)\right\}\bm{U_{\pi^S}}(T, R, S; \hat{\pi}^S).
\end{align*}

\section*{Appendix B: Estimator for Known Selection Model}
In this section we focus on the locally semi-parametric efficient estimator, which is more efficient compared to the proposed estimator but not doubly robust.

\subsection*{Appendix B.1: Theoretical Derivation of the Estimator}
Following similar reasonings as in Shu and Tan \cite{ShuTan2018}, and assuming that the selection model is known, $\theta_1$ can be estimated as 
\begin{align*}
\hat{\theta}_{1}^{eff}
%&=\frac{\sum_{i=1}^{n} \frac{I(R_i=f)T_i}{\hat{\pi}^f\hat{\pi}_T}(Y_i-h(\bm{Z}_i, \bm{\hat{\eta}}))+\frac{\pi(\bm{Z}; \hat{\bm{\gamma}})}{\hat{\pi}^T}h(\bm{Z}_i, \bm{\hat{\eta}})}{\sum_{i=1}^{n} \frac{\pi(\bm{Z}; \hat{\bm{\gamma}})}{\hat{\pi}_T}}\\
&=\frac{n^{-1}\sum_{i=1}^{n} \frac{I(R_i=f)T_i}{\hat{\pi}^f}(Y_i-h(\bm{Z}_i, \bm{\hat{\eta}}))+\pi(\bm{Z}; \hat{\bm{\gamma}})h(\bm{Z}_i, \bm{\hat{\eta}})}{n^{-1}\sum_{i=1}^{n} \pi(\bm{Z}; \hat{\bm{\gamma}})}.
\end{align*}
Using a logistic selection model for $\pi(\bm{Z}; \bm{\gamma})$, this reduces to 
\begin{align*}
\hat{\theta}_{1}^{eff'}
&=\frac{n^{-1}\sum_{i=1}^{n} \frac{I(R_i=f)T_i}{\hat{\pi}^f}(Y_i-h(\bm{Z}_i, \bm{\hat{\eta}}))+\pi(\bm{Z}; \hat{\bm{\gamma}})h(\bm{Z}_i, \bm{\hat{\eta}})}{\hat{\pi}^T}.
\end{align*}
By estimating $\hat{\bm{\eta}}$ as the solution to the equations $\sum_{i=1}^n\frac{\text{I}(R_i=f)}{\hat{\pi}^f}(1, \bm{Z}_i)'\left(Y_i-h(\bm{Z}_i;\bm{\eta})\right)$, it follows that the estimator $\hat{\theta}_{1}^{eff'}$ reduces to 
$n^{-1}\sum_{i=1}^{n}\pi(\bm{Z}; \hat{\bm{\gamma}})h(\bm{Z}_i, \bm{\hat{\eta}})/\hat{\pi}^T$; which coincides with the semi-parametric efficient estimator proposed for $\theta_1$ in the main paper. 

A similar reasoning for  $\theta_2$ yields
\begin{align*}
\begin{split}
\hat{\theta}_{2}^{eff}=&\frac{n^{-1}\sum_{i=1}^n\frac{\text{I}(T_i=0)\text{I}(R_i=l)}{(1-\pi(\bm{Z}_i,\hat{\bm{\gamma}}))\hat{\pi}^l}\pi(\bm{Z}_i,\hat{\bm{\gamma}})\left(Y_i-m(\bm{Z}_i;\hat{\bm{\beta}})\right)+\pi(\bm{Z}_i,\hat{\bm{\gamma}})m(\bm{Z}_i;\hat{\bm{\beta}})\vphantom{\frac{P(S=1|X)}{P(S=0|X)}} }{n^{-1}\sum_{i=1}^{n} \pi(\bm{Z}; \hat{\bm{\gamma}})}.
\end{split}
\end{align*}
Using a logistic selection model, $\pi(\bm{Z}; \bm{\gamma})$, this reduces to 
\begin{align*}
\hat{\theta}_{2}^{eff'}
&=\frac{n^{-1}\sum_{i=1}^n\frac{\text{I}(T_i=0)\text{I}(R_i=l)}{(1-\pi(\bm{Z}_i,\hat{\bm{\gamma}}))\hat{\pi}^l}\pi(\bm{Z}_i,\hat{\bm{\gamma}})\left(Y_i-m(\bm{Z}_i;\hat{\bm{\beta}})\right)+\pi(\bm{Z}_i,\hat{\bm{\gamma}})m(\bm{Z}_i;\hat{\bm{\beta}})\vphantom{\frac{P(S=1|X)}{P(S=0|X)}} }{\hat{\pi}^T}.
\end{align*}
By estimating $\hat{\bm{\beta}}$ as the solution to the equations $\sum_{i=1}^n\frac{\text{I}(T_i=0)\text{I}(R_i=l)}{(1-\pi(\bm{Z}_i,\hat{\bm{\gamma}}))\hat{\pi}^l}\pi(\bm{Z}_i,\hat{\bm{\gamma}})(1, \bm{Z}_i)'\left(Y_i-m(\bm{Z}_i;\bm{\beta})\right)$, it follows that the estimator $\hat{\theta}_{2}^{eff'}$ reduces to 
$\sum_{i=1}^n\left(\pi(\bm{Z}_i,\hat{\bm{\gamma}})m(\bm{Z}_i;\hat{\bm{\beta}})\right)/\hat{\pi}^T$; which coincides with the semi-parametric effictient estimator proposed for $\theta_2$ in the main paper. 

The variance of these estimators can be obtained by following a similar reasoning as in Appendix A.3.

\subsection*{Appendix B.2: Remark on the (Double) Robustness of the Estimator}\label{app_remark}
Because of simple randomisation, $\hat{\pi}^f$ and $\hat{\pi}^l$ are consistent estimator for $P(R=f|T=1)$ and $P(R=l|T=0)$. Additionally, $\hat{\pi}^T$ is a consistent estimator for $P(T=1)$. Define the probability limits $\bm{\eta^*}=\text{plim}(\bm{\hat{\eta}})$, $\bm{\beta^*}=\text{plim}(\bm{\hat{\beta}})$ and $\bm{\gamma^*}=\text{plim}(\bm{\hat{\gamma}})$, which equal respectively the true values $\bm{\eta}_0$,  $\bm{\beta}_0$ and $\bm{\gamma}_0$ when the working models $h(\bm{Z}, \bm{\eta})$, $m(\bm{Z}, \bm{\beta})$ and $\pi(\bm{Z}, \bm{\gamma})$ are correctly specified, but not necessarily otherwise. 

By the weak law of large numbers
and Slutsky's theorem, $\hat{\theta}_2^{eff}$ estimates
\begin{align}\label{est_asympt}
\begin{split}
E\left[\frac{\text{I}(T=0)\text{I}(R=l)\pi(\bm{Z};\bm{\gamma^*})}{\left\{1-\pi(\bm{Z};\bm{\gamma^*})\right\}\pi^l}\left\{Y-m(\bm{Z};\bm{\beta^*})\right\}+\pi(\bm{X};\bm{\gamma^*})m(\bm{Z};\bm{\beta^*})\right]\Big/E\left[\pi(\bm{Z};\bm{\gamma^*})\right].
\end{split}
\end{align}
First, consider the case where the working model for the selection, $\pi(\bm{Z};\bm{\gamma})$, is correctly specified and the workig model for the outcome, $m(\bm{Z};\bm{\beta})$, possibly incorrectly specified. Expression (\ref{est_asympt}) then reduces to,
\begin{align*}
&E\left[\frac{\text{I}(T=0)\text{I}(R=l)\pi(\bm{Z};\bm{\gamma_0})}{\left\{1-\pi(\bm{Z};\bm{\gamma_0})\right\}\pi^l}\left\{Y-m(\bm{Z};\bm{\beta^*})\right\}+\pi(\bm{X};\bm{\gamma_0})m(\bm{Z};\bm{\beta^*})\right]\Big/E\left[\pi(\bm{Z};\bm{\gamma_0})\right]\\
&=E\left[\left(\pi(\bm{Z};\bm{\gamma_0})-\frac{\text{I}(T=0)\text{I}(R=l)\pi(\bm{Z};\bm{\gamma_0})}{\left\{1-\pi(\bm{Z};\bm{\gamma_0})\right\}\pi^l}\right)m(\bm{Z};\bm{\beta^*})+\frac{\text{I}(T=0)\text{I}(R=l)\pi(\bm{Z};\bm{\gamma_0})}{\left\{1-\pi(\bm{Z};\bm{\gamma_0})\right\}\pi^l}Y\right]\Big/E\left[T\right]\\
&=E\left[\frac{\text{I}(T=0)\text{I}(R=l)\pi(\bm{Z};\bm{\gamma_0})}{\left\{1-\pi(\bm{Z};\bm{\gamma_0})\right\}\pi^l}Y\right]\Big/E\left[T\right]\\
&=E\left[\pi(\bm{Z};\bm{\gamma_0})E(Y|T=0, R=l, \bm{Z})\right]\Big/E\left[T\right]\\
&=E\left[\pi(\bm{Z};\bm{\gamma_0})E(Y^l|T=0, \bm{Z})\right]\Big/E\left[T\right]\\
&=E\left[\pi(\bm{Z};\bm{\gamma_0})E(Y^l|T=1, \bm{Z})\right]\Big/E\left[T\right]\\
&=E\left[ E(TY^{l}|\bm{Z})\right]/E[T]\\
&=E\left[ TY^{l}\right]/E[T]\\
&=E\left[ Y^{l}|T=1\right].
\end{align*}
This estimator, however, has the disadvantage not to be robust against model misspecification of $\pi(\bm{Z};\bm{\gamma})$. To see that, consider the case where the working model for the outcome, $m(\bm{Z};\bm{\beta})$, is correctly specified and the workig model for the propensity score, $\pi(\bm{Z};\bm{\gamma})$, possibly incorrectly specified. Expression (\ref{est_asympt}) then reduces to,
\begin{align*}
&E\left[\frac{\text{I}(T=0)\text{I}(R=l)\pi(\bm{Z};\bm{\gamma^*})}{\left\{1-\pi(\bm{Z};\bm{\gamma^*})\right\}\pi^l}\left\{Y-m(\bm{Z};\bm{\beta_0})\right\}+\pi(\bm{X};\bm{\gamma^*})m(\bm{Z};\bm{\beta_0})\right]\Big/E\left[\pi(\bm{Z};\bm{\gamma^*})\right]\\
&=E\left[\pi(\bm{Z};\bm{\gamma^*})m(\bm{Z};\bm{\beta_0})\right]/E[\pi(\bm{Z};\bm{\gamma^*})],
%&=E\left[\pi(\bm{Z};\bm{\gamma^*})Y^{l}\right]/E[\pi(\bm{Z};\bm{\gamma^*})],
\end{align*}
where the equation follows from the way $\bm{\beta}$ is estimated. This expectation only leads to $E[Y^{l}|T=1]$ if the covariates used to estimate $m(\bm{Z};\bm{\beta})$ are a subset of the covariates used to estimate $\pi(\bm{Z};\bm{\gamma})$. Then, by the estimating equation of a logistic regression with an intercept $E\left[\pi(\bm{Z};\bm{\gamma^*})m(\bm{Z};\bm{\beta_0})\right]=E\left[T\cdot m(\bm{Z};\bm{\beta_0})\right]$ and $E[\pi(\bm{Z};\bm{\gamma^*})]=E[T]$. Moreover,
\begin{align*}
E\left[T\cdot m(\bm{Z};\bm{\beta_0})\right]/E[T]
&=E\left[T\cdot E(Y|T=0, R=l, \bm{Z})\right]/E[T]\\
&=E\left[T\cdot E(Y^{l}|T=0, \bm{Z})\right]/E[T]\\
&=E\left[T\cdot E(Y^{l}|T=1, \bm{Z})\right]/E[T]\\
&=E\left[P(T=1|\bm{Z}) E(Y^{l}|T=1, \bm{Z})\right]/E[T]\\
&=E\left[ E(TY^{l}|\bm{Z})\right]/E[T]\\
&=E\left[ TY^{l}\right]/E[T]\\
&=E\left[ Y^{l}|T=1\right].
\end{align*}
Consequently, both estimators for $\theta_2$ ($\hat{\theta}_2$ and $\hat{\theta}_2^{eff}$) are equivalent when using a logistic regression for $\pi(\bm{Z}, \bm{\gamma})$ with a set of covariates that includes the covariates used in the model $m(\bm{Z};\bm{\beta})$. In that case, $\hat{\theta}_2^{eff}$ is doubly robust.
A similar reasoning for $\theta_1$ shows that both estimators for $\theta_1$ ($\hat{\theta}_1$ and $\hat{\theta}_1^{eff}$) are equivalent when using a logistic regression for $\pi(\bm{Z}, \bm{\gamma})$ with a set of covariates that includes the covariates used in the model $h(\bm{Z};\bm{\eta})$.
To conclude, both estimators for $\theta$ are equivalent when using a logistic regression for $\pi(\bm{Z}, \bm{\gamma})$ with a set of covariates that includes the covariates used in the models $m(\bm{Z};\bm{\beta})$ and $h(\bm{Z};\bm{\eta})$.

\section*{Appendix C: Tables}
%\begin{figure}
%	\centering
%	\begin{subfigure}{0.6\textwidth}
%		\centering
%		\includegraphics[width=\linewidth]{Baseline_fixed1}
%		%		\caption{A subfigure}
%		%		\label{fig:sub1}
%	\end{subfigure}%
%	\vspace{0.1cm}
%	\begin{subfigure}{0.6\textwidth}
%		\centering
%		\includegraphics[width=\linewidth]{Baseline_fixed2}
%		%		\caption{A subfigure}
%		%		\label{fig:sub2}
%	\end{subfigure}
%	\caption{Baseline and demographic characteristics: fixed dosing study.}
%	\label{fig:baseline_fixed}
%\end{figure}
%\begin{figure}
%	\centering
%	\begin{subfigure}{0.6\textwidth}
%		\centering
%		\includegraphics[width=\linewidth]{Baseline_flex1}
%		%		\caption{A subfigure}
%		%		\label{fig:sub1}
%	\end{subfigure}%\\
%	\vspace{0.1cm}
%	\begin{subfigure}{0.6\textwidth}
%		\centering
%		\includegraphics[width=\linewidth]{Baseline_flex2}
%		%		\caption{A subfigure}
%		%		\label{fig:sub2}
%	\end{subfigure}
%	\caption{Baseline and demographic characteristics: flexible dosing study.}
%	\label{fig:baseline_flex}
%\end{figure}

\begin{table}
	\centering		
	\begin{threeparttable}
		\caption{\label{tab: results_large}Comparison of the bias, the standard error and the mean squared error of the regression estimator and the estimators proposed in Section \ref{sec:proposal} for a total sample size of $5000$, under a misspecified outcome model.}
		\begin{tabular}{l l c ccc c ccc c ccc}
			\hline
			 &  & & \multicolumn{3}{c}{Non-parametric} && \multicolumn{3}{c}{Semi-parametric} && \multicolumn{3}{c}{Regression}\\ \cline{4-6}
			\cline{8-10} \cline{12-14}
			 Setting & Weights&&Bias&SE&MSE&&Bias&SE&MSE&&Bias&SE&MSE\\ \hline
			Setting $1$ &$1.85; 2.11$ &&%$4.252\cdot 10^{-5}$ 
			$0$
			& $0.023$&$0.023$&&$-0.001$& $0.022$&$0.022$&&$-0.001$& $0.024$ &$0.024$\\
			Setting $2$ & $0.85; 3.80$ &&$0.008$& $0.026$&$0.026$&&$0.007$& $0.025$& $0.025$&&$0.711$ &$0.026$ &$0.531$\\
			Setting $3$ & $0.34; 6.13$ &&$0.028$& $0.039$&$0.040$&&$0.026$& $0.039$ & $0.039$&&$1.527$ &$0.030$ & $2.360$\\
			Setting $4$ & $0.03; 6.63$ &&$0.343$& $0.212$&$0.330$&&$0.342$& $0.212$&$0.329$&&$3.054$& $0.048$&$9.377$\\
			Setting $5$ & $0.001; 3.36$ &&$1.396$& $0.691$&$2.638$&&$1.394$& $0.690$&$2.633$&&$4.086$& $0.085$ & $16.777$\\
			\hline
			\hline
		\end{tabular}
\begin{tablenotes}
	\small
	\item Note: The column weights shows the $5\%$ and $95\%$ percentiles of the weights $\hat{\pi}(\bm{Z}_i, \hat{\bm{\gamma}})/(1-\hat{\pi}(\bm{Z}_i, \hat{\bm{\gamma}}))$ among the patients on the low dose of the fixed dosing trial.. 
\end{tablenotes}
	\end{threeparttable}
\end{table}

\begin{table}[h]
	\centering		
	\begin{threeparttable}
		\caption{\label{tab: results_violated}Comparison of the bias, the standard error and the mean squared error of the regression estimator and the estimators proposed in Section \ref{sec:proposal} under violation of the mean exchangeability assumption.}
		\begin{tabular}{l l c ccc c ccc c ccc}
	\hline
	&  & & \multicolumn{3}{c}{Non-parametric} && \multicolumn{3}{c}{Semi-parametric} && \multicolumn{3}{c}{Regression}\\ \cline{4-6}
	\cline{8-10} \cline{12-14}
	Setting & Weights&&Bias&SE&MSE&&Bias&SE&MSE&&Bias&SE&MSE\\ \hline
			Setting $1$ & $1.77; 2.21$ &&$-0.001$ &$0.147$&$0.147$&&$-0.002$& $0.121$& $0.121$&&$-0.001$& $0.147$ & $0.147$\\
			Setting $2$ & $1.08; 3.10$ &&$0.471$& $0.164$&$0.385$&&$0.470$& $0.137$&$0.358$&&$0.470$& $0.159$&$0.380$\\
			Setting $3$ & $0.58; 3.89$ &&$0.941$ &$0.212$&$1.098$&&$0.941$ &$0.188$&$1.073$&&$0.940$& $0.186$&$1.070$\\
			Setting $4$ & $0.10; 4.23$ &&$1.873$ &$0.592$&$4.098$&&$1.873$& $0.566$&$4.072$&&$1.880$& $0.321$&$3.854$\\
			Setting $5$ & $0.02; 3.66$ &&$2.819$& $1.585$&$9.532$&&$2.821$& $1.559$&$9.515$&&$2.820$& $0.584$&$8.535$\\
			\hline
			\hline
		\end{tabular}
	\begin{tablenotes}
		\small
		\item Note: The column weights shows the $5\%$ and $95\%$ percentiles of the weights $\hat{\pi}(\bm{Z}_i, \hat{\bm{\gamma}})/(1-\hat{\pi}(\bm{Z}_i, \hat{\bm{\gamma}}))$ among the patients on the low dose of the fixed dosing trial.. 
	\end{tablenotes}
		%	\begin{tablenotes}
		%		\small
		%	\item Note: NP, eff. estimator under non-parametric model; SP, eff. estimator under under semi-parametric model. 
		%\end{tablenotes}
	\end{threeparttable}
\end{table}

\begin{landscape}
\begin{table}[h]
	\centering		
	\begin{threeparttable}
		\caption{\label{tab: results_violated2}Comparison of the bias, the standard error and the mean squared error of the regression estimator and the estimators proposed in Section \ref{sec:proposal} under violation of the mean exchangeability assumption, measuring $X_{11}$ instead of $X_7$.}
		\begin{tabular}{ll c c ccc c ccc c ccc}
	\hline
	&  &  & & \multicolumn{3}{c}{Non-parametric} && \multicolumn{3}{c}{Semi-parametric} && \multicolumn{3}{c}{Regression}\\ \cline{5-7} \cline{9-11} \cline{13-15}
	Correlation & Setting & Weights&&Bias&SE&MSE&&Bias&SE&MSE&&Bias&SE&MSE\\ \hline
			$0.8$ & Setting $1$ & $1.84; 2.52$ &&$-0.008$ &$0.123$&$0.123$&&$-0.005$& $0.099$&$0.099$&&$-0.008$& $0.123$ & $0.123$\\
			& Setting $2$ & $1.16; 3.44$ &&$0.088$ &$0.133$&$0.141$&&$0.090$& $0.112$&$0.120$&&$0.089$ &$0.131$&$0.139$\\
			& Setting $3$ & $0.49; 5.72$ &&$0.184$& $0.172$&$0.206$&&$0.185$& $0.153$&$0.187$&&$0.186$& $0.154$&$0.188$\\
			& Setting $4$ & $0.05; 4.25$ &&$0.379$& $0.558$&$0.701$&&$0.378$ &$0.539$&$0.682$&&$0.377$& $0.302$&$0.445$\\
			& Setting $5$ & $0.002; 3.22$ &&$0.565$& $3.393$&$3.712$&&$0.564$& $3.354$&$3.672$&&$0.559$& $1.444$&$1.756$\\
			\hline
			$0.5$ & Setting $1$ & $1.83; 2.57$ &&$-0.008$& $0.137$&$0.137$&&$-0.005$& $0.113$&$0.113$&&$-0.008$ &$0.137$&$0.137$\\
			& Setting $2$ & $1.19; 3.34$ &&$0.230$& $0.156$&$0.209$&&$0.232$ &$0.133$&$0.187$&&$0.231$& $0.151$&$0.204$\\
			& Setting $3$ & $0.50; 5.00$ &&$0.469$& $0.216$&$0.436$&&$0.469$& $0.195$&$0.415$&&$0.470$ &$0.187$&$0.407$\\
			& Setting $4$ & $0.05; 4.04$ &&$0.941$& $0.822$&$1.708$&&$0.940$& $0.798$&$1.681$&&$0.942$ &$0.419$&$1.307$\\
			& Setting $5$ & $0.002; 3.51$ &&$1.403$& $4.266$&$6.233$&&$1.402$& $4.222$&$6.186$&&$1.409$& $1.794$&$3.779$\\
			\hline
			\hline
		\end{tabular}
	\begin{tablenotes}
		\small
		\item Note: The column weights shows the $5\%$ and $95\%$ percentiles of the weights $\hat{\pi}(\bm{Z}_i, \hat{\bm{\gamma}})/(1-\hat{\pi}(\bm{Z}_i, \hat{\bm{\gamma}}))$ among the patients on the low dose of the fixed dosing trial.. 
	\end{tablenotes}
		%	\begin{tablenotes}
		%		\small
		%	\item Note: NP, eff. estimator under non-parametric model; SP, eff. estimator under under semi-parametric model. 
		%\end{tablenotes}
	\end{threeparttable}
\end{table}
\end{landscape}

% Table created by stargazer v.5.2.2 by Marek Hlavac, Harvard University. E-mail: hlavac at fas.harvard.edu
% Date and time: Tue, Jul 30, 2019 - 12:03:42
% Requires LaTeX packages: dcolumn 
\begin{table}[!htbp] \centering 
	\caption{Regression Results for $h(\bm{Z}, \bm{\eta})$. Covariates: COUNTRY, country (reference: Czech Republic); AGE, age at baseline (centered); WEIGHTBL, weight at baseline (centered); YBL, baseline measurement of the outcome $Y$ (centered).} 
	\label{tab:regressionY2} 
	\begin{tabular}{@{\extracolsep{5pt}}lD{.}{.}{-3} } 
		\\[-1.8ex]\hline 
		\hline \\[-1.8ex] 
		& \multicolumn{1}{c}{\textit{Dependent variable: $Y$}} \\ 
		\cline{2-2} 
		\\[-1.8ex] Predictors & \multicolumn{1}{c}{Coefficient (Standard Error)} \\ 
		\hline \\[-1.8ex] 
		Constant & -24.140 (2.076)^{***} \\
		COUNTRY&\\
		\hspace{0.5cm}GERMANY & 13.690 (4.200)^{***} \\  
		\hspace{0.5cm}SPAIN & -18.292 (8.186)^{**} \\ 
		\hspace{0.5cm}POLAND & 4.219 (3.370) \\ 
		\hspace{0.5cm}USA & 4.476 (2.724) \\ 
		AGE & 0.171 (0.182) \\ 
		WEIGHTBL & -0.140 (0.058)^{**} \\ 
		YBL & -1.147 (0.197)^{***} \\ 
		COUNTRY:AGE& \\ 
		\hspace{0.5cm}GERMANY:AGE & 0.222 (0.404) \\  
		\hspace{0.5cm}SPAIN:AGE & 2.160 (0.690)^{***} \\ 
		\hspace{0.5cm}POLAND:AGE & -0.085 (0.274) \\ 
		\hspace{0.5cm}USA:AGE & -0.094  (0.222) \\ 
		WEIGHTBL:YBL & 0.013 (0.009) \\  
		AGE:WEIGHTBL & -0.006 (0.004) \\ 
		\hline \\[-1.8ex] 
		Observations & \multicolumn{1}{c}{112} \\ 
		R$^{2}$ & \multicolumn{1}{c}{0.410} \\ 
		Adjusted R$^{2}$ & \multicolumn{1}{c}{0.332} \\ 
		Residual Std. Error & \multicolumn{1}{c}{11.105 (df = 98)} \\ 
		F Statistic & \multicolumn{1}{c}{5.238$^{***}$ (df = 13; 98)} \\ 
		\hline 
		\hline \\[-1.8ex] 
		\textit{Note:}  & \multicolumn{1}{r}{$^{*}$p$<$0.1; $^{**}$p$<$0.05; $^{***}$p$<$0.01} \\ 
	\end{tabular} 
\end{table}

% Table created by stargazer v.5.2.2 by Marek Hlavac, Harvard University. E-mail: hlavac at fas.harvard.edu
% Date and time: Tue, Jul 30, 2019 - 10:57:23
% Requires LaTeX packages: dcolumn 
\begin{table}[h!] \centering 
	\caption{Regression Results for $m(\bm{Z}, \bm{\beta})$. Covariates: YBL, baseline measurement of the outcome $Y$ (centered); HEIGHTBL, height at baseline (centered); ETHNIC, ethnicity (reference: hispanic or latino).} 
	\label{tab:regressionY} 
	\begin{tabular}{@{\extracolsep{5pt}}lD{.}{.}{-3} } 
		\\[-1.8ex]\hline 
		\hline \\[-1.8ex] 
				& \multicolumn{1}{c}{\textit{Dependent variable: $Y$}} \\ 
		\cline{2-2} 
		\\[-1.8ex] Predictors & \multicolumn{1}{c}{Coefficient (Standard Error)} \\ 
		\hline \\[-1.8ex] 
		Constant & -25.534 (2.601)^{***}  \\ 
		YBL & 0.200 (0.563) \\ 
		HEIGHTBL & -0.259 (0.230) \\ 
		ETHNIC&\\ 
		\hspace{0.5cm}NOT HISPANIC OR LATINO & 9.659 (3.022)^{***}  \\ 
		\hspace{0.5cm}UNKNOWN & 10.625 (7.665) \\ 
		YBL:HEIGHTBL & 0.165 (0.069)^{**}  \\ 
		YBL:ETHNIC&\\
		\hspace{0.5cm}YBL:NOT HISPANIC OR LATINO & -1.372 (0.652)^{**}  \\ 
		\hspace{0.5cm}YBL:UNKNOWN & 2.176  (1.913) \\ 
		HEIGHTBL:ETHNIC&\\
		\hspace{0.5cm}HEIGHTBL:NOT HISPANIC OR LATINO & 0.032 (0.289) \\ 
		\hspace{0.5cm}HEIGHTBL:UNKNOWN & 0.523 (0.915) \\
		YBL:HEIGHTBL:ETHNIC&\\ 
		\hspace{0.5cm}YBL:HEIGHTBL:NOT HISPANIC OR LATINO & -0.200 (0.080)^{**}  \\  
		\hspace{0.5cm}YBL:HEIGHTBL:UNKNOWN & 0.049 (0.243) \\ 
		\hline \\[-1.8ex] 
		Observations & \multicolumn{1}{c}{115} \\ 
		R$^{2}$ & \multicolumn{1}{c}{0.251} \\ 
		Adjusted R$^{2}$ & \multicolumn{1}{c}{0.171} \\ 
		Residual Std. Error & \multicolumn{1}{c}{12.939 (df = 103)} \\ 
		F Statistic & \multicolumn{1}{c}{3.144$^{***}$ (df = 11; 103)} \\ 
		\hline 
		\hline \\[-1.8ex] 
		\textit{Note:}  & \multicolumn{1}{r}{$^{*}$p$<$0.1; $^{**}$p$<$0.05; $^{***}$p$<$0.01} \\ 
	\end{tabular} 
\end{table}

% Table created by stargazer v.5.2.2 by Marek Hlavac, Harvard University. E-mail: hlavac at fas.harvard.edu
% Date and time: Tue, Jul 30, 2019 - 11:40:37
% Requires LaTeX packages: dcolumn 
\begin{table}[!htbp] \centering 
	\caption{Regression Results for $\pi(\bm{Z}, \bm{\gamma})$. Covariate: ETHNIC, ethnicity.} 
	\label{tab:regressionSelection} 
	\begin{tabular}{@{\extracolsep{5pt}}lD{.}{.}{-3} } 
		\\[-1.8ex]\hline 
		\hline \\[-1.8ex] 
				& \multicolumn{1}{c}{\textit{Dependent variable: $T$}} \\ 
		\cline{2-2} 
		\\[-1.8ex] Predictors & \multicolumn{1}{c}{Coefficient (Standard Error)} \\ 
		\hline \\[-1.8ex] 
		Constant & -2.026 (0.307)^{***} \\ 
		ETHNIC&\\
		\hspace{0.5cm}NOT HISPANIC OR LATINO & 1.942 (0.322)^{***} \\ 
		\hspace{0.5cm} UNKNOWN & 0.116  (0.618) \\ 
		\hline \\[-1.8ex] 
		Observations & \multicolumn{1}{c}{562} \\ 
		Log Likelihood & \multicolumn{1}{c}{-345.279} \\ 
		Akaike Inf. Crit. & \multicolumn{1}{c}{696.558} \\ 
		\hline 
		\hline \\[-1.8ex] 
		\textit{Note:}  & \multicolumn{1}{r}{$^{*}$p$<$0.1; $^{**}$p$<$0.05; $^{***}$p$<$0.01} \\ 
	\end{tabular} 
\end{table} 

\begin{landscape}
	\begin{table}[h]
		\centering	
		\resizebox{\columnwidth}{!}{%	
			\begin{threeparttable}
				\caption{\label{tab: sensitivity} Treatment estimator, $95\%$ confidence interval and $p$-value for the treatment difference obtained in the primary analysis and the sensitivity analyses.}
				\begin{tabular}{ll ccc}
					\hline
					&&\multicolumn{3}{c}{Selection Model}\\\cline{3-5}
					Outcome Model & & ETHNIC & ETHNIC+RACE+REGION+PRIOR & ETHNIC$*$RACE+REGION+PRIOR\\ \hline
					ETHNIC$*$YBL$*$HEIGHTBL & & $-4.311$ $[-9.404, 0.782]$ & $-5.458$ $[-10.906, -0.010]$ & $-5.838^*$ $[-11.122, -0.553]$\\
					& & $p=0.096$ & $p=0.049$ & $p=0.030$\\
					&&&&\\
					ETHNIC+YBL+HEIGHTBL & & $-4.832$ $[-9.795, 0.131]$ & $-6.152$ $[-11.558, -0.746]$ & $-5.944$ $[-11.334, -0.553]$\\
					& & $p=0.056$ & $p=0.025$ & $p=0.030$\\
					&&&&\\
					ETHNIC$*$YBL$*$HEIGHTBL+ & & / & $-7.069$ $[-12.999, -1.139]$ & $-7.364^{**}$ $[-13.240, -1.487]$\\
					REGION+PRIOR+RACE& & / & $p=0.019$ & $p=0.014$\\
					&&&&\\
					ETHNIC+YBL+HEIGHTBL+ & & / & $-7.718$ $[-13.742, -1.693]$ & $-7.503$ $[-13.527, -1.478]$\\
					REGION+PRIOR+RACE& & / & $p=0.012$ & $p=0.014$\\
					\hline \hline
				\end{tabular}
				\begin{tablenotes}
					\small
					\item Note: the categories "unknown" and "North-America" of the variable REGION are merged to one categorie for computational reasons.\\
					$^*$ Due to computational problems as a consequence of categories with only a few individuals, the outcome model was replaced by I(KNOWN)+I(HISPANIC OR LATINO)$*$YBL$*$HEIGHTBL; where I(KNOWN) equals $1$ if ETHNIC equals ``HISPANIC OR LATINO'' or ``NOT HISPANIC OR LATINO'' and $0$ if it equals ``UNKNOWN'', and I(HISPANIC OR LATINO) equals $1$ if ETHNIC equals ``HISPANIC OR LATINO'' and $0$ otherwise.\\
					$^{**}$ Due to computational problems as a consequence of categories with only a few individuals, the outcome model was replaced by I(KNOWN)+I(HISPANIC OR LATINO)$*$YBL$*$HEIGHTBL+REGION+%OLANTDMD
					PRIOR+RACE\\
					Note: ETHNIC, ethnicity; YBL, baseline measurement of the outcome $Y$;  HEIGHTBL, height at baseline; REGION, region; %OLANTDMD, oral antidepressant use prior to randomization; 
					RACE, race; PRIOR, treatment initiated prior to randomization.
				\end{tablenotes}
			\end{threeparttable}
		}
	\end{table}
\end{landscape}

\begin{figure}
	\centering
		\centering
		\includegraphics[width=0.75\linewidth]{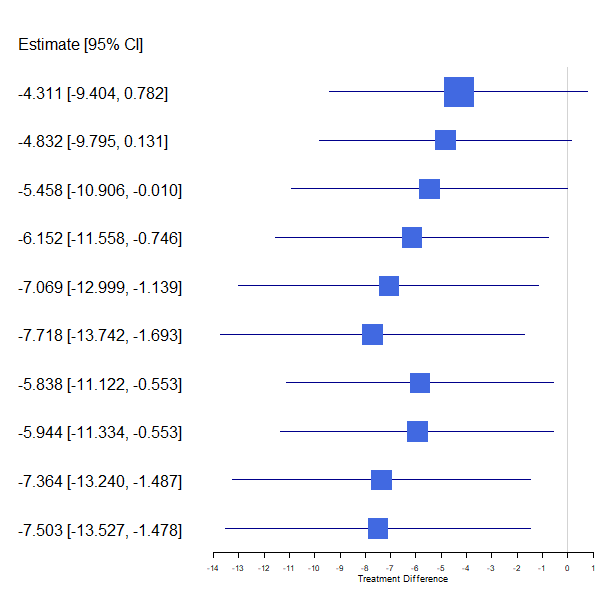}
	\caption{Forest plot for the treatment difference obtained in the primary analysis (big square) and the sensitivity analyses.}
	\label{fig:ForestPlot}
\end{figure}

\newpage

\bibliographystyle{Chicago} %Chicago
%https://ajp.psychiatryonline.org/doi/pdf/10.1176/appi.ajp.2008.08071102

%https://www.ncbi.nlm.nih.gov/pmc/articles/PMC3308934/
%https://www.ncbi.nlm.nih.gov/pmc/articles/PMC1140678/
%https://books.google.be/books?id=ETIFX7HyiowC&pg=PA134&lpg=PA134&dq=%22two+positive+studies%22+FDA&source=bl&ots=FQQWKUGUeB&sig=ACfU3U2bQGUoa8REdzFeri8C3-qCo4Twag&hl=nl&sa=X&ved=2ahUKEwjztsum667iAhVqSBUIHZQGDE04ChDoATAEegQICBAB#v=onepage&q=%22two%20positive%20studies%22%20FDA&f=false

\bibliography{biblio_correct3}

\end{document}